\documentclass[aps,amssymb,amsfonts,pra,superscriptaddress,twocolumn, floatfix, notitlepage]{revtex4-1}
\usepackage[utf8]{inputenc}
\usepackage{amsmath}
\usepackage{graphicx}
\usepackage[usenames]{color}
\usepackage{amssymb}
\usepackage{supertabular}
\usepackage{subfigure}
\usepackage{physics}
\usepackage{soul}
\usepackage{multirow}
\usepackage[english]{babel}
\usepackage{comment}
\usepackage[T1]{fontenc}
\usepackage[ruled,vlined,linesnumbered]{algorithm2e}
\usepackage[export]{adjustbox}

\newcommand{\beq}{\begin{equation}}
\newcommand{\eeq}{\end{equation}}

\def\ra{{\rangle}}

\newcommand{\ie}{\textit{i.e.}, }

\newcommand{\A}{\mathcal{A}}
\newcommand{\D}{\mathcal{D}}
\newcommand{\C}{\mathcal{C}}
\newcommand{\R}{\mathcal{R}}
\newcommand{\Q}{\mathcal{Q}}
\renewcommand{\S}{\mathcal{S}}

\begin{document}
\title{Quantum unary approach to option pricing}

\author{Sergi Ramos-Calderer}
\email{sergi.ramos@tii.ae}
\affiliation{Departament de F\'isica Qu\`antica i Astrof\'isica and Institut de Ci\`encies del Cosmos (ICCUB), Universitat de Barcelona, Mart\'i i Franqu\`es 1, 08028 Barcelona, Spain.}
\affiliation{Quantum Research Centre, Technology Innovation Institute, Abu Dhabi, UAE.}

\author{Adrián P\'{e}rez-Salinas}
\affiliation{Departament de F\'isica Qu\`antica i Astrof\'isica and Institut de Ci\`encies del Cosmos (ICCUB), Universitat de Barcelona, Mart\'i i Franqu\`es 1, 08028 Barcelona, Spain.}
\affiliation{Barcelona Supercomputing Center (BSC), Spain.}

\author{Diego García-Martín}
\affiliation{Departament de F\'isica Qu\`antica i Astrof\'isica and Institut de Ci\`encies del Cosmos (ICCUB), Universitat de Barcelona, Mart\'i i Franqu\`es 1, 08028 Barcelona, Spain.}
\affiliation{Barcelona Supercomputing Center (BSC), Spain.}
\affiliation{Instituto de Física Teórica, UAM-CSIC, Madrid, Spain.}

\author{Carlos Bravo-Prieto}
\affiliation{Departament de F\'isica Qu\`antica i Astrof\'isica and Institut de Ci\`encies del Cosmos (ICCUB), Universitat de Barcelona, Mart\'i i Franqu\`es 1, 08028 Barcelona, Spain.}
\affiliation{Barcelona Supercomputing Center (BSC), Spain.}

\author{\\Jorge Cortada}
\affiliation{Caixabank, Barcelona, Spain.}

\author{Jordi Planagum\`a}
\affiliation{Caixabank, Barcelona, Spain.}

\author{Jos\'{e} I. Latorre}
\affiliation{Departament de F\'isica Qu\`antica i Astrof\'isica and Institut de Ci\`encies del Cosmos (ICCUB), Universitat de Barcelona, Mart\'i i Franqu\`es 1, 08028 Barcelona, Spain.}
\affiliation{Quantum Research Centre, Technology Innovation Institute, Abu Dhabi, UAE.}
\affiliation{Center for Quantum Technologies, National University of Singapore, Singapore.}

\begin{abstract}
We present a quantum algorithm for European option pricing in finance, where the key idea is to work in the unary representation of the asset value. The algorithm needs novel circuitry and is divided in three parts: first, the amplitude distribution corresponding to the asset value at maturity is generated using a low depth circuit; second, the computation of the expected return is computed with simple controlled gates; and third, standard Amplitude Estimation is used to gain quantum advantage. On the positive side, unary representation remarkably simplifies the structure and depth of the quantum circuit. Amplitude distributions uses quantum superposition to bypass the role of classical Monte Carlo simulation. The unary representation also provides a post-selection consistency check that allows for a substantial mitigation in the error of the computation. On the negative side, unary representation requires linearly many qubits to represent a target probability distribution, as compared to the logarithmic scaling of binary algorithms. We compare the performance of both unary {\sl vs.} binary option pricing algorithms using error maps, and find that unary representation may bring a relevant advantage in practice for near-term devices.
\end{abstract}

\maketitle
\section{Introduction}

Quantum computing provides new strategies to address problems that nowadays are considered difficult to solve by classical means. The first quantum algorithms showing a theoretical advantage over their classical counterparts are known since the 1990s, such as integer factorization to prime numbers \cite{factorization-shor1999} or a more efficient unstructured database search \cite{search-grover1997}.
Nevertheless, current quantum devices are not powerful enough to run quantum algorithms that are able to compete against state-of-the-art classical algorithms. Indeed, available quantum computers are in their Noisy Intermediate-Scale Quantum (NISQ) stage \cite{nisq-preskill2018}, as errors due to decoherence, noisy gate application or error read-out limit the performance of these new machines. These NISQ devices may nonetheless be useful tools for a variety of applications due to the introduction of hybrid variational methods. Some of the proposed applications include quantum chemistry \cite{vqe-peruzzo2014, vqe-higgott2019, vqe-jones2019}, simulation of physical systems \cite{vqs-li2017, vqs-kokail2019, vqs-cirstoiu2019}, combinatorial optimization \cite{qaoa-fahri2014}, solving large systems of linear equations \cite{vls-bravo2019, vls-xu2019, vls-huang2019}, state diagonalization \cite{qsd-larose2019, qsd-bravo2019} or quantum machine learning \cite{qml-mitarai2018, qml-zhu2019, qml-perez2019}. Some exact, non-variational, quantum algorithms are also well suited for NISQ devices \cite{ising-cervera2018, entanglement-subasi2018, quantum-bravyi2018, quantum-bravyi2019}. 

A field that is expected to be transformed by the improvement of quantum devices is quantitative finance \cite{qfinance-orus2019, portfolio-kerenidis2019, crashes-orus2019, derivatives-martin2019, credit-egger2019}. In recent years, there has been a surge of new methods and algorithms dealing with financial problems using quantum resources, such as optimization problems \cite{optimization-rosenberg2016, optimization-rebentrost2018, optimization-moll2018, optimization-lopez2015} which are in general hard.

Notably, pricing of financial derivatives is a prominent problem, where many of its computational obstacles are suited to be overcome via quantum computation. In this paper we will deal with options, which are a particular type of financial derivatives. Options are contracts that allow the holder to buy (\textit{call}) or sell (\textit{put}) some asset at a pre-established price (\textit{strike}), or at a future point in time (\textit{maturity date}). The payoff of an option depends on the evolution of the asset's price, which follows a stochastic process. A simple, yet successful model for pricing options is the Black-Scholes model \cite{blackscholes-black1973}. This is an analytically-solvable model that predicts the asset's price evolution to follow a log-normal probability distribution, at a future time $t$. Then, a specified payoff function, which depends on the particular option considered, has to be integrated over this distribution to obtain the expected return of the option. Current classical algorithms rely on computationally-costly Monte Carlo simulations to estimate the expected return of options.

A few quantum algorithms have been proposed to improve on classical option pricing \cite{qfinance-stamatopoulos2019, qfinance-rebentrost2018, qfinance-woerner2019}. It has been shown that quantum computers can provide a quadratic speedup in the number of quantum circuit runs as compared to the number of classical Monte Carlo runs needed to reach a certain precision in the estimation. The basic idea is to exploit quantum Amplitude Estimation \cite{amplitude_estimation-brassard2002, counting-aaronson2019, montecarlo-montanaro2015quantum}. Nonetheless, this can only be achieved when an efficient way of loading the probability distribution of the asset price is available. The idea of using quantum Generative Adversarial Networks (qGANs) \cite{qGAN-lloyd2018, qGAN-dallaire2018} to address this issue has been analyzed \cite{qGAN-zoufal2019}.

In the following, we propose a quantum algorithm for option pricing. The key new idea is to construct a quantum circuit that works in the unary basis of the asset's value, \ie in a subspace of the full Hilbert space of $n$ qubits. Then, the evolution of the asset's price is computed using an amplitude distributor module. Furthermore, the computation of the payoff greatly simplifies. A third part of the algorithm is common to previous approaches, namely it uses Amplitude Estimation. The unary scheme brings further advantage since it allows for a post-selection strategy that results in error mitigation. Let us recall that error mitigation techniques are likely to be crucial for the success of quantum algorithms in the NISQ era. On the negative side, the number of qubits in the unary algorithm scales linearly with the number of bins, while in the binary algorithm it is logarithmic with the target precision. This results in a worse asymptotic scaling for the unary algorithm. Yet, our estimates for the number of gates indicate that the crossing point between these two is located at a number of qubits that renders a good precision ($< 1\%$) for real-world applications. Moreover, the performance of the unary algorithm is more robust to noise, as we show in simulations. Hence, our proposal seems to be better suited to be run on NISQ devices. Unary representations have also been considered in previous works \cite{spectral-poulin2018, babbush2018,steudtner2019}.

We will illustrate our new algorithm focusing on a simple European option, whose payoff is a function of only the asset's price at maturity date, the only date the contract can be executed at. This straightforward example has been chosen as a proof of concept for this new approach. We will compare the performance of our unary quantum circuit with the previous binary quantum circuit proposal, for a fixed precision or binning of the probability distribution.

The paper is organized as follows. We first introduce the basic ideas on option pricing, both classical and quantum, in Sec. \ref{sec:background}. The unary quantum algorithm is presented and analyzed in Sec. \ref{sec:unary}. We devote Sec. \ref{sec:un-vs-bin} to outline the circuit specifications and compare them for the unary and binary quantum algorithms. Sec. \ref{sec:simulations} is dedicated to describe the results obtained by means of classical simulations for both algorithms. Lastly, conclusions are drawn in Sec. \ref{sec:conclusions}. Further details on several topics are described in the Appendices.

\section{Background}\label{sec:background}

There are three main pieces that lay the groundwork needed for our algorithm. They are a) the economical model employed in European-option pricing, known as the Black-Scholes model; b) the Amplitude Estimation technique that provides a quadratic quantum advantage over classical Monte Carlo methods; and c) a quantum algorithm for option pricing in the binary basis, as proposed in \cite{qfinance-stamatopoulos2019}.

\subsection{Black-Scholes model}\label{sec:econ_model}

The evolution of asset prices in financial markets is usually computed using the model established by F. Black and M. Scholes in Ref. \cite{blackscholes-black1973}. This evolution is governed by two properties of the market, the interest rate and the volatility, which are incorporated into a stochastic differential equation.

The Black-Scholes model for the evolution of an asset's price at time $T$, $S_T$, is based on the following stochastic differential equation,
\begin{equation}\label{eq:BSM}
    {\rm d}S_T = S_T\, r\, {\rm d}T + S_T\, \sigma\, {\rm d}W_T, 
\end{equation}
where $r$ is the interest rate, $\sigma$ is the volatility and $W_T$ describes a Brownian process.
Let us recall that a Brownian process $W_T$ is a continuous stochastic evolution starting at $W_0=0$ and consisting of independent gaussian increments. To be specific, let $\mathcal{N}(\mu, \sigma_s)$ be a normal distribution with mean $\mu$ and standard deviation $\sigma_s$. Then, the increment related to two steps of the Brownian processes is  $W_T - W_S \sim \mathcal{N}(0, T - S)$, for $T > S$.

The stochastic differential equation \eqref{eq:BSM} can be approximately solved analytically to first order, yielding the solution 
\begin{equation}\label{eq:log_normal}
    S_T = S_0 e^{(r - \frac{\sigma^2}{2}) T} e^{\sigma W_T}\;\sim\; e^{\mathcal{N}\left(\left(r - \frac{\sigma^2}{2}\right) T, \sigma \sqrt{T}\right)},
\end{equation}
which corresponds to a log-normal distribution. The details of this procedure are outlined in App. \ref{sec:ap_econ_model}.

To obtain the expected return of an option, a payoff function has to be integrated over the
resulting probability distribution. This is usually solved using classical Monte Carlo simulation.

In the case of European options, the payoff function is
\begin{equation}
    f(S_T, K) = \max(0, S_T - K),
\end{equation}
yielding an expected payoff given by

\begin{equation}\label{eq:avg_payoff}
    C(S_T, K) = \int_K^\infty \left( S_T - K \right)\,dS_T,
\end{equation}
where $K$ is the strike. European options can only be executed at a fixed pre-specified time, called {\sl maturity date}. This is the reason why the payoff is computed using only the probability distribution of $S_T$ at time $T$.

Our algorithm employs a quantum circuit that generates a probability distribution following Eq. \eqref{eq:log_normal}, and then encodes the expected payoff of a European option, Eq. \eqref{eq:avg_payoff}, into the amplitudes of an ancilla qubit.

\subsection{Amplitude Estimation\label{sec:AE}}

Amplitude Estimation (AE) is a quantum technique that allows to estimate the probability of obtaining a certain outcome from a quantum state (with a given precision), with up to a quadratic speedup in the number of function calls as compared to direct sampling \cite{amplitude_estimation-brassard2002, amplitude_estimation-suzuki2020}.

\subsubsection*{AE with Quantum Phase Estimation}
Let us take an algorithm $\A$ such that \begin{equation}
    \A \ket 0_n \ket 0 = \sqrt{1 - a} \ket{\psi_0}_n\ket{0} + \sqrt{a} \ket{\psi_1}_n\ket{1}, 
\end{equation}
where the last qubit serves as an ancilla qubit and the states $\ket{\psi_{0,1}}_n$ can be non-orthogonal. The ancilla qubit is a flag which enables to identify the states as {\sl good} ($\ket{1}$) or {\sl bad} ($\ket{0}$). The state $\A\ket 0_n \ket 0$ can be directly sampled $N$ times, and the estimate for probability of finding a good outcome will be $\bar a$, with
\begin{equation}
    |a - \bar a| \sim \mathcal{O}(N^{-1/2}),
\end{equation}
as dictated by the sampling error of a multinomial distribution.

However, AE can improve this result. Let us first define the central operator for AE \cite{amplitude_estimation-brassard2002}
\begin{equation}
    \Q = - \A \S_0 \A^\dagger \S_{\psi_0},
\end{equation}
where the operators $\S_0$ and $\S_{\psi_0}$ are inherited from Grover's search algorithm \cite{search-grover1997}, being
\begin{eqnarray}
    \S_0 & = & \mathbf{I} - 2 \ket 0_n \bra 0_n \otimes \ket 0 \bra 0 , \\ 
    \S_{\psi_0} & = & \mathbf{I} - 2 \ket{\psi_0}_n\bra{\psi_0}_n \otimes \ket 0 \bra 0.
\end{eqnarray}
The $\S_0$ operator changes the sign of the $\ket 0_n \ket 0$ state, while $\S_{\psi_0}$ takes the role of an oracle and changes the sign of all bad outcomes. The operator $\Q$ has eigenvalues $e^{\pm i 2 \theta_a}$, with $a = \sin^2(\theta_a)$. The procedure of Quantum Phase Estimation (QPE) is then applied to extract an integer number $y \in \{0, 1, \ldots, 2^m-1\}$ such that $\bar{\theta}_a = \pi y / 2^m$ is an estimate of $\theta_a$, with $m$ the number of ancilla qubits. Recall that a Quantum Fourier Transform (QFT) is required to perform QPE.

The value of $\bar{\theta}_a$ leads to an estimate of $\bar a$, such that
\begin{equation}
  |a - \bar a| < \frac{2\pi \sqrt{a (1 - a)}}{2^m} + \frac{\pi^2}{2^{2m}} \sim \mathcal{O}\left(\frac{\pi}{2^m}\right)  
\end{equation}
with probability at least $8/\pi^2\approx 81\%$. 

The original Amplitude Estimation procedure requires the implementation of QPE, which is highly resource demanding. Hence, the complexity of the circuit precludes its feasibility in the NISQ era.

\subsubsection*{AE without Quantum Phase Estimation}
Recently, there has been a new proposal for Amplitude Estimation that does not require QPE, and therefore, is less resource-demanding. This approach is based on iterative procedures \cite{amplitude_estimation-suzuki2020}. The key fact allowing to circumvent the use of QPE is that
\begin{equation}\label{eq:q_j}
\begin{split}
    \Q^{m} \A \ket 0 = \cos\left((2 m + 1)\theta_a\right)\ket{\psi_0}_n \ket{0} + \\ + \sin\left((2 m + 1)\theta_a\right)\ket{\psi_1}_n \ket{1}.
\end{split}
\end{equation}
An integer $m$ is chosen to prepare the state in Eq. \eqref{eq:q_j} and its outcome is measured $N_{\rm shots}$ times, so that the value of $\sin^2\left((2 m + 1)\theta_a\right)$ is estimated with a precision of $\sim N_{\rm shots}^{-1/2}$. This process is repeated several times with different values of $m$ extracted from a set of $\{m_j\}$. At the end of the procedure, the precision achieved is bounded by $\sim N_{\rm shots}^{-1/2}M^{-1}$, with $M=\sum_{j=0}^J m_j$, where $J$ is the last index. The exact scaling of the precision depends on the choice of $m_j$'s. In App. \ref{sec:ap_iae} the full method is explained in further detail.

\subsection{Binary algorithm}\label{sec:binary}

We now present a binary algorithm for option pricing, as introduced in Ref. \cite{qfinance-stamatopoulos2019}. This algorithm is divided in three parts. 
\begin{itemize}
    \item[(a)] {\sl Amplitude distributor}: it encodes the underlying probability distribution of an asset's price at maturity date into a quantum register. The operator representing this piece will be denoted by $\D$. This algorithm uses a quantum Generative Adversarial Network (qGAN) \cite{qGAN-lloyd2018, qGAN-dallaire2018, qGAN-zoufal2019} in order to fulfill this part. Classical knowledge of the probability distribution is required at this stage.
    \item[(b)] {\sl Payoff calculation}: it computes the expected payoff of the option, which is encoded into the amplitude of an ancillary qubit. The operators that perform this step will be a comparator $\C$, that separates the state as above or below the strike, and a set of controlled rotations $\R$, that encode the expected payoff into the probability of measuring an ancilla.
    \item[(c)] {\sl Amplitude Estimation}: it extracts the expected payoff calculation encoded in the amplitude of the ancilla, reducing the number of circuit calls needed to reach a desired precision. It is based on the operator $\Q$, which may be applied several times.
\end{itemize}

\begin{figure}
    \centering
\includegraphics[width=\linewidth]{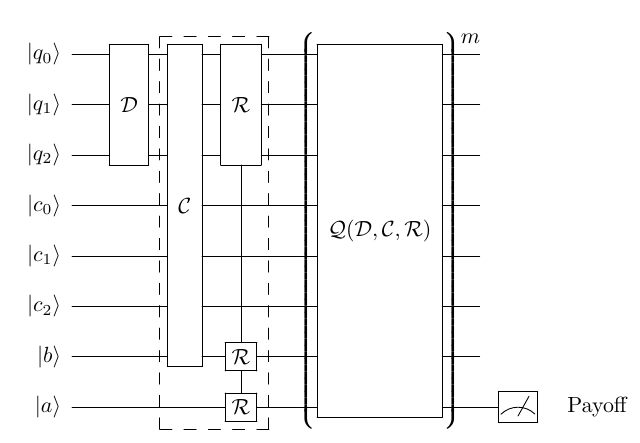}
    \caption{Full circuit for the binary algorithm for option pricing that include all steps, namely, the amplitude distributor $\D$, payoff estimator comprised of the comparator and payoff estimator $\C$ and $\R$ respectively, followed by components of Amplitude Estimation, $\Q$. The operator $\Q$ is repeated $m$ times, where $m$ depends on the AE algorithm. The payoff is indirectly measured in the last qubit.}
    \label{fig:binary_circuit}
\end{figure}

A sketch of a quantum circuit implementing the full algorithm is shown in Fig. \ref{fig:binary_circuit}. For a detailed description of each part, refer to App. \ref{sec:ap_binary}.

\begin{figure*}[t!]
    \centering
    \includegraphics[width=.7\linewidth]{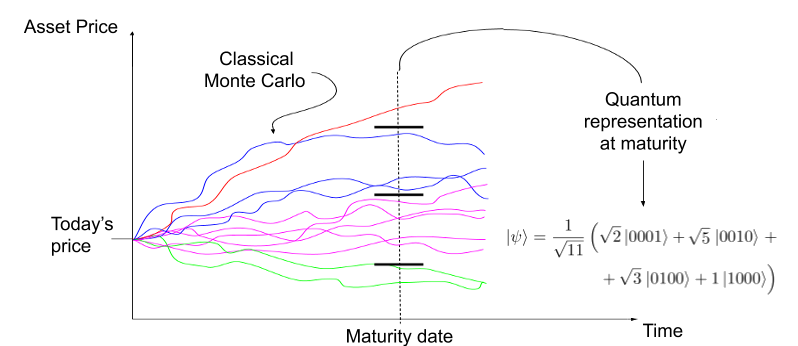}
    \caption{Scheme for the quantum representation of a given asset price at maturity date. For a given number of Monte Carlo paths, a binning scheme must be applied in such a way that the prices of the asset are separated according to its value. Different Monte Carlo paths that end up in the same bin are color coded accordingly. Each bin is mapped then to an element of the unary basis, whose coefficient is the number of Monte Carlo paths in this bin.
    The quantum representation of the asset price at maturity contains all possible Monte Carlo paths simultaneously. The precision is then  bounded by the numbers of bins we can store on a quantum state, \ie how many qubits are available.}
    \label{fig:MC}
\end{figure*}

\section{Unary algorithm}\label{sec:unary}

We present now a quantum algorithm that prices European options according to the Black-Scholes model, as outlined in Sec. \ref{sec:econ_model}. The key new idea is to construct a quantum circuit that works in the unary basis of the asset's value. The structure of the algorithm is inherited from the one explained in Sec. \ref{sec:binary}, namely: amplitude distributor module, computation of the payoff and Amplitude Estimation. Furthermore, the implementation of all different pieces is greatly simplified with respect to the binary case. The unary scheme brings further advantage in practice, since it allows for a post-selection strategy that results in error mitigation. Although our unary algorithm requires more qubits than a binary one, its performance is more robust to noise and probably better suited to be run on NISQ devices. 

\subsection{Unary representation}
The main feature of the algorithm is that it works in the {\sl unary representation} of the asset value encoded on the quantum register. That means that for every element of the basis only one qubit will be in the $\ket 1$ state, whereas all others will remain in $\ket 0$. A quantum register $\ket \psi_n$ made of $n$ qubits in the unary representation can be written as
\begin{equation}
    \begin{split}
    \ket\psi &= \sum_{i=0}^{n-1} \psi_i \ket{i}_n = \sum_{i=0}^{n-1} \psi_i \left(\bigotimes_{j=0}^{n-1} \ket{\delta_{i j}}\right) = \\
    &= \psi_0 \ket{00\ldots 01}_n+\psi_1 \ket{00\ldots 10}_n+\ldots \\ 
    & \qquad+ \psi_{n-2} \ket{01\ldots 00}_n+\psi_{n-1} \ket{10\ldots 00}_n ,
    \end{split}
\end{equation}
where $\ket i_n$ corresponds to the $i$-th element of the unary basis, $\delta_{i j}$ is the Kronecker delta and $\sum_{i=1}^n |\psi_i|^2=1$. A well-known example of a state in the unary representation is the $W$ state, which defines a class of three-qubit multipartite entanglement \cite{entanglement-dur2000}. Depicted in Fig. \ref{fig:MC} is a visual representation of how the unary algorithm would map the outcomes of a Monte Carlo simulation of the asset's price to a quantum register. The ratio of Monte Carlo paths leading to each of the bins will translate into the amplitudes of the corresponding unary basis states.

Given a fixed number of qubits, the unary scheme allows for a lower precision than the binary one. Indeed, only $n$ out of the $2^n$ basis elements of the Hilbert space are used. However, due to the natural mapping between the unary representation and the asset's price evolution, we will find that the probability distribution loading and the expected payoff calculation can be carried out with much simpler quantum circuits. On real devices, the potential gain of the unary representation translates into a shallower circuit depth and simpler connectivity requirements. Furthermore, the unary scheme provides means to post-select results so as to increase the faithfulness of the computation. This is due to the fact that the unary representation resides within a restricted part of the Hilbert space, and that extra space can be used as an indicator of the appearance of errors. As a matter of fact, given a realistic precision goal (<1\%), it may well turn out to be advantageous to move to the unary representation on NISQ devices, as it simplifies the complexity of the circuit and mitigates errors. 

In most cases, a quantum computation does not start in a quantum state that belongs to the unary representation, but in the $\ket 0_n$ state instead. To solve this issue, we act with a single Pauli X-gate on any qubit. In our algorithm, this qubit is chosen to be the central one to improve overall circuit depth. At this point, the register displays a single qubit in $\ket{1}$ while all others are in $\ket 0$. This register is an element of the unary basis. 

\subsection{Implementation of the algorithm \label{sec:implementation_unary}}

The basic structure of the unary algorithm is directly inherited from the structure of the binary one. All three independent parts are {\sl Amplitude distributor}, {\sl Payoff calculator} and {\sl Amplitude Estimation}. We discuss them now in further detail.

\subsubsection*{Amplitude distributor}

The probability distribution predicted by the Black-Scholes model is based on the one in Eq. \eqref{eq:log_normal}.
For a given number of qubits, that is of precision, the asset price at any time can be mapped to a fixed depth circuit that distributes probabilities according to the final desired result.
The unary representation is akin to the value of the asset. In other words, for every element in the superposition describing the quantum register, the qubit which is flipped into $\ket{1}$ determines the value of the asset.
The classical Monte Carlo spread of asset values will be mapped into the probability of measuring each unary basis element.

The quantum circuit generating the final register operates as a distributor of probability amplitudes. 
The initial state of the algorithm is given by $\ket{0\ldots 010\ldots 0}_n$, \ie the element of the unary basis with $\ket 1$ in the middle qubit. Then, the coefficients of the register in the next step of the circuit are generated using partial-SWAP gates (also called parametrized-SWAP or SWAP power gate) between the middle qubit and its neighbors. The partial-SWAP gate is defined as 
\begin{equation}\label{eq:SWAPRy}
      \includegraphics[width=0.3\linewidth, valign=c]{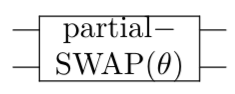} =
    \left(\begin{array}{cccc}
    1&0&0&0 \\ 0&\cos{\theta/2}&\sin{\theta/2}&0 \\0&-\sin{\theta/2}&\cos{\theta/2}&0 \\0&0&0&1
    \end{array}\right)
\end{equation}
Moreover, the partial-SWAP gate could be substituted with a partial-iSWAP gate which performs the same purpose of amplitude sharing. This partial-iSWAP gate,
\begin{equation}\label{eq:p-iSWAP}
    \includegraphics[width=0.3\linewidth, valign=c]{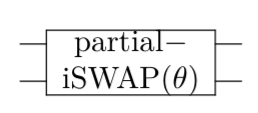} =\left(\begin{array}{cccc}
    1&0&0&0 \\ 0&\cos{\theta/2}&-i\sin{\theta/2}&0 \\0&-i\sin{\theta/2}&\cos{\theta/2}&0 \\0&0&0&1
    \end{array}\right),
\end{equation}
is a universal entangling gate that comes naturally from the capacitive coupling of superconducting qubits \cite{partialiSWAP-bialczak2010, iSWAP-schuch2003}. As a matter of fact, Google's Sycamore chip in which the supremacy experiment was performed \cite{supremacy2019} allows for this type of gates as they are of great importance as well for quantum chemistry applications \cite{barkoutsos2018quantum, gard2020efficient} or combinatorial optimization \cite{hadfield2019qaoa, wang2019xy, cook2019xy}.

\begin{figure}[ht]
    \includegraphics[width=\linewidth]{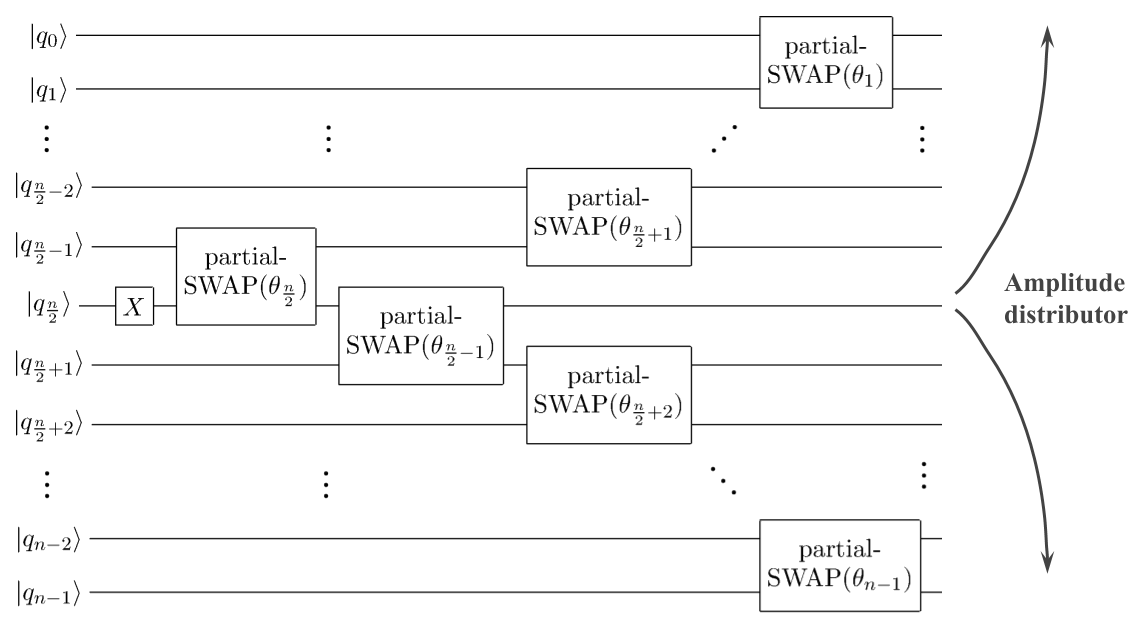}
\caption{Quantum circuit for loading any probability distribution in the unary representation $\D$. The circuit works as a distributor of amplitude probabilities from its middle qubit to the ones in the edges, using partial-SWAP gates that act only on nearest neighbors. Time dependence is encoded in the angles determining the gates.}
\label{fig:ProbUploading}
\end{figure}

This provides the first step to distribute the probability amplitude from the middle qubit to the rest. The procedure is repeated until the edge of the system is reached, as illustrated in Fig. \ref{fig:ProbUploading}. Specific angles can be fed into each partial-SWAP gate to obtain the target probability distribution in the unary representation. The detailed procurement of these angles is described in App. \ref{sec:ap_amplitude_distribution}.

Let us note that any final probability distribution at time $t$ can be obtained with this circuit whose depth is independent of time, since all the necessary information is carried in the angles of the partial-SWAP gates. To be precise, given $n$ qubits, the circuit will always be of depth $\lfloor{n/2}\rfloor+1$. The time dependency of the solution is encoded in the angles determining the partial-SWAP gates. This idea is reminiscent of the quantum circuits that describe the exact solution of the Ising model \cite{verstraete2009,hebenstreit2017,ising-cervera2018}.

The mapping of a known probability distribution function to the unary system is dependant on $(n-1)$ angles that need to be introduced in the partial-SWAP gates. There are two distinct situations depending on whether the final distribution probability is known exactly or not. The first case can be addressed solving an exact set of $n$ equations with $n-1$ parameters after computing the probability distribution classically. 
In case only the differential equation is known, but not its solution, other methods should be employed \cite{diffeq-iblisdir2007}. 

\subsubsection*{Payoff calculator}

The expected payoff calculation circuit builds upon the action of the amplitude distributor to encode the expected return on an ancillary qubit. The unary representation allows for a simple algorithm to accomplish this task. The procedure will prepare an entangled state in the form
\begin{equation}\label{eq:AEstate}
\ket{\Psi}=\sqrt{1-a}\ket{\psi_0}_n\ket{0} +
    \sqrt{a}\ket{\psi_1}_n\ket{1},
\end{equation}
where $\ket{\psi_{0,1}}$ are states in a superposition of the basis elements below and above the strike, respectively.
The payoff is encoded within the amplitude $\sqrt{a}$, with $|a| \leq 1$, ready for Amplitude Estimation \cite{montecarlo-montanaro2015quantum}.

\begin{figure}[t!]
    \centering
    \minipage{0.5\textwidth}
      \includegraphics[width=\linewidth]{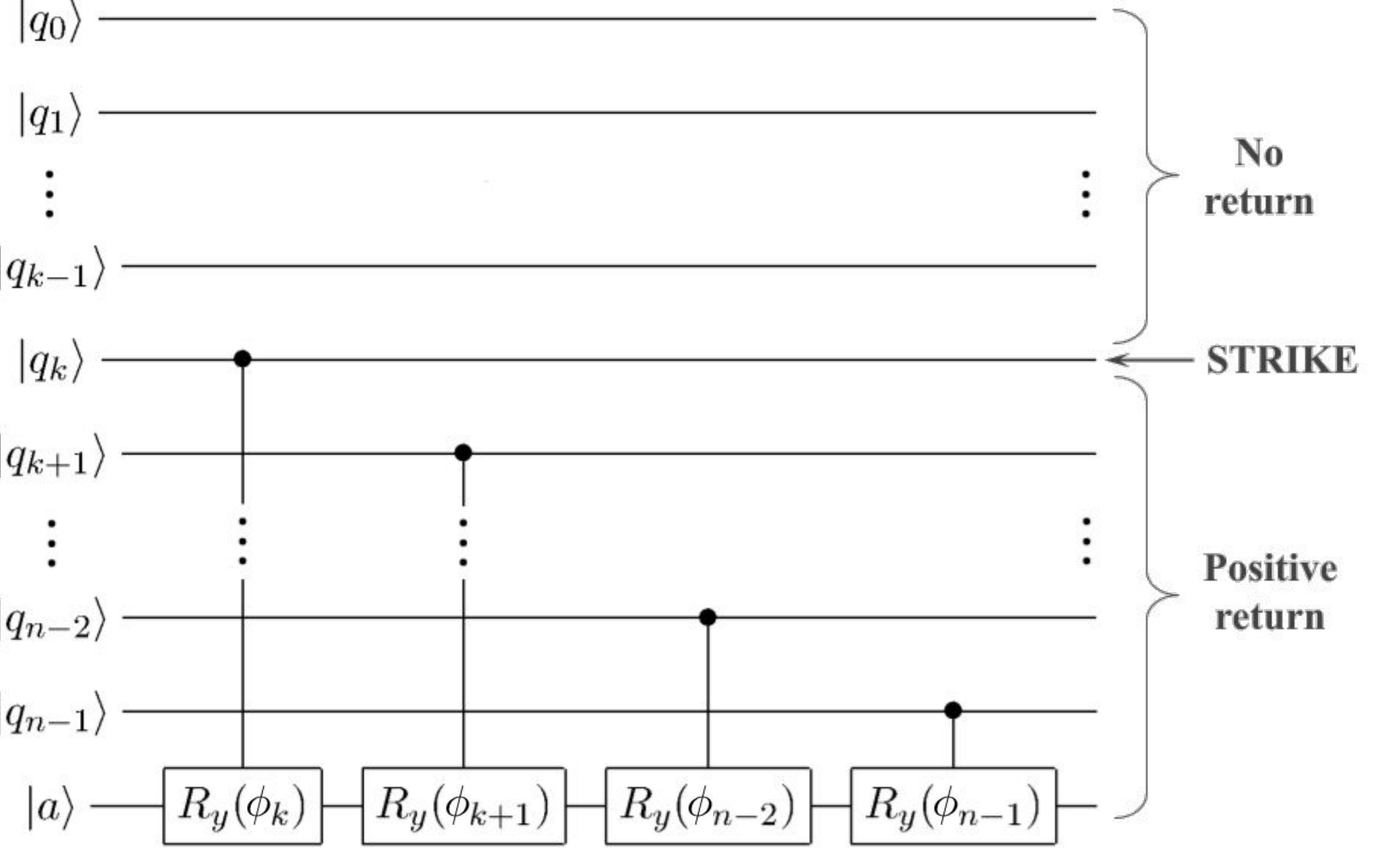}
    \endminipage\hfill
\caption{Quantum circuit that encodes the expected payoff in an ancillary qubit in the unary representation $\C + \R$. Each qubit with a mapped option price higher than the designated strike controls a c$R_y$ gate on the ancilla, where the rotation angle is a function of its contribution to the expected payoff. The comparator $\C$ is constructed through the control wires, while the $\R$ piece is performed by rotations in the last qubit.}
\label{fig:PayoffCircuit}
\end{figure}

The relevant point to encode the payoff of a European option in an ancillary qubit is to distinguish in the quantum register whether the option price $S_i$ is above or below the strike $K$. This task turns out to be very simple when working in the unary representation, as opposed to the binary one where a comparator $\C$ needs to be introduced. To be explicit, the computation of the payoff can be achieved by applying controlled Y rotations (c$R_y$ gates), whose control qubits are those encoding
a price higher than the accorded strike $K$, namely the operator $\R$.
These c$R_y$ gates will only span over those qubits that represent asset values larger than the strike. Note that the depth of the circuit will be $n-k$, where $k$ is the unary label of the strike $K$, see Fig. \ref{fig:PayoffCircuit}.

The rotation angle for each c$R_y$ depends on the contribution of the qubit to the expected payoff. This can be achieved using
\begin{equation}
    \phi_i=2\arcsin\sqrt{\frac{S_i-K}{S_{max}-K}},
\end{equation}
where the denominator inside the $\arcsin$ argument is introduced for normalization. 

Applying the payoff calculator to a quantum state representing the probability distribution, as depicted in Fig. \ref{fig:MC}, results in 
\begin{equation}
\begin{split}
    \ket{\Psi}=\sum_{S_i\leq K}^{n-1}\sqrt{p_i}\ket{i}_n\ket{0}+\sum_{S_i>K}^{n-1}\sqrt{p_i}\cos(\phi_i/2)\ket{i}_n\ket{0}+\\
    +\sum_{S_i>K}^{n-1}\sqrt{p_i}\sqrt{\frac{S_i-K}{S_{max}-K}}\ket{i}_n\ket{1}.
\end{split}
\end{equation}
The state is now in the form of Eq. \eqref{eq:AEstate}. It is straightforward to see that the probability of measuring $\ket 1$ in the ancillary qubit is
\begin{equation}
    P(\ket 1) = \sum_{S_i > K} p_i \frac{S_i - K}{S_{max} - K}.
\end{equation}
In order to recover the encoded expected payoff, we need to measure the probability of obtaining $\ket{1}$ for the ancilla and then multiply it by the normalization factor ${S_{max}-K}$. Note that the form of the state is such that further Amplitude Estimation can be performed. 

\subsubsection*{Amplitude Estimation}

\begin{figure}[t]
    \centering
    \hspace{1cm} 
    \includegraphics[width=.4\linewidth]{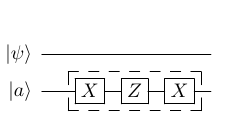}
    \hfill 
    {
    \includegraphics[width=.4\linewidth]{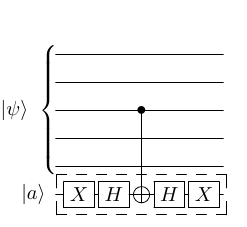}}\hspace{1cm}
    \caption{Quantum circuit representation of $\S_{\phi_0}$ (Left) and $\S_0$  (Right) required to perform Amplitude Estimation in the unary basis. Notice that operator $\S_0$ is much simpler in the unary representation as it does not require multi-controlled CNOT gates.}
    \label{fig:ae_circuits}
\end{figure}

Let us now move to the application of Amplitude Estimation to our unary option pricing algorithm. As described in Sec. \ref{sec:AE}, Amplitude Estimation is performed by concatenating the operators $\A$ and its inverse $\A^\dagger$ with operators $\S_0$ and $\S_{\psi_0}$. In the following, we will describe how to implement these operators in the unary algorithm. Detailed implementation can be seen as well in Fig. \ref{fig:ae_circuits}.

\begin{figure*}
\includegraphics[width=\linewidth]{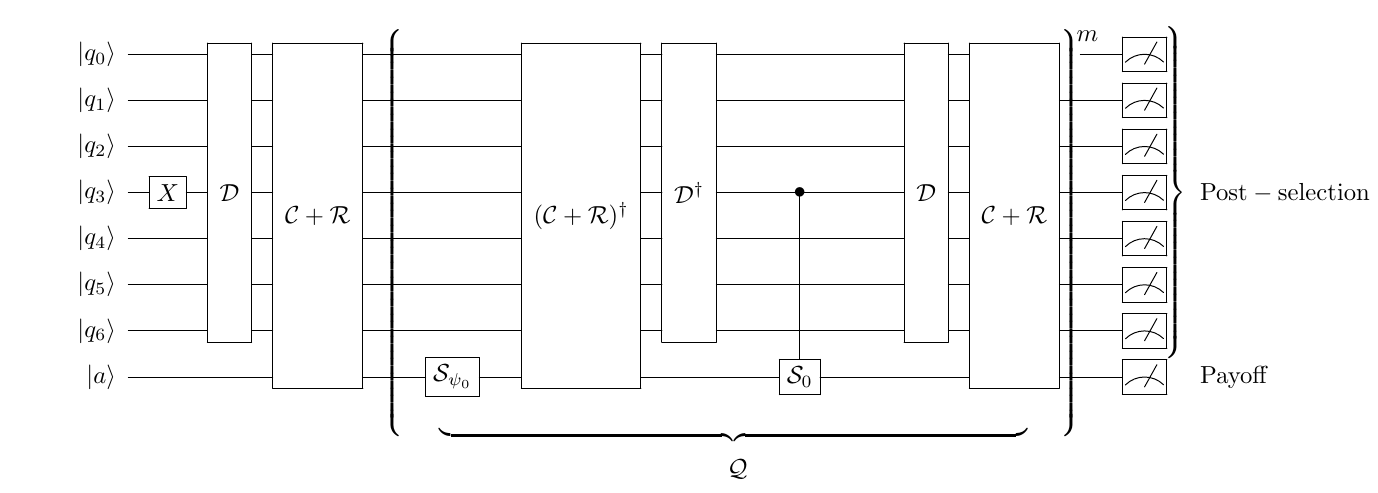}
    \caption{Full circuit for the option pricing algorithm in the unary representation. The gate $\D$ is the probability distributor, and $\C + \R$ represent the computation of the payoff. After applying the algorithm, the oracle $\S_{\psi_0}$, the reverse algorithm and $\S_0$ follow. The last step is applying the algorithm again. This block $\Q$ is to be repeated for Amplitude Estimation. Measurements in all qubits is a requirement for post-selection. The qubit labelled as $q_3$ is the one starting the unary representation. }
    \label{fig:full_unary}
\end{figure*}

The oracle operator $\S_{\psi_0}$ acts by identifying those coefficients corresponding to accepted outcomes and inverting their sign. In this problem, the task of identifying the element of the basis has been already done by the algorithm $\A$. Accepted outcomes are labelled with $\ket 1$ in the ancilla qubit. Therefore, the function of this oracle can be achieved by local operations in the ancilla qubit. Explicitly, such operation is
\begin{equation}\label{eq:oracle}
    \S_{\psi_0} = (I^{\otimes n} \otimes (XZX)).
\end{equation}
Notice that the $X$-gates can be deleted since they only provide a global sign. 

For the case of the operator $\S_0$ we must remark a detail that greatly simplifies this computation.
The operator $\S_0$ is normally defined using $\ket{0}$ since most quantum algorithms start on that state, as depicted in Eq. \eqref{eq:AEstate}. However, a more apt definition should instead include $\ket{initial}$ as a basis for operator $\S_0$, the state onto which the algorithm $\A$ is first applied. For the unary case, if we isolate the first extra $X$ gate, we can consider the algorithm as starting in that state of the unary basis, heavily simplifying the overall construction. That being the case, $\S_0$ can be constructed out of 2 single-qubit gates and one entangling gate.

With the operator $\Q$ constructed, Amplitude Estimation schemes can be performed. Since the unary algorithm is aimed towards NISQ devices, we use an Amplitude Estimation scheme without Quantum Phase Estimation, explained in detail in App. \ref{sec:ap_iae}. The main idea consists in applying operator $\Q$ a different amount of $m$ times, and process the data in order to get an advantage over ordinary sampling.

\subsection{Error mitigation\label{subsec:error_unary}} 

NISQ era algorithms need to be resilient against gate errors and decoherence, since fault-tolerant logical qubits are still far from being a reality. Error mitigation techniques have been studied in past literature, see Refs. \cite{temme-mitigation2017, endo-mitigation2018}, and some of them might find valid applications in the unary algorithms as well. However, the unary representation we are proposing here turns out to offer an additional, native, post-selection strategy that manages to mitigate different types of errors. This feature is not present in its binary counterpart.

The key idea behind the possibility of accomplishing error mitigation is that unary algorithms should ideally work within the unary subspace of the Hilbert space. As a consequence, the read-out of any measurement should reflect this fact. It is then possible to reject any outcome that does not fulfil this requirement. As a matter of fact, a number of failed repetitions of the experiment could be discarded, what results in a trade-off between reduction of errors and loss of accepted samples. 

A scheme for the full circuit is depicted in Fig. \ref{fig:full_unary}. In summary, the circuit is composed by one first $X$ gate that initializes the unary basis, one set of amplitude distributor $(\D)$ and payoff calculator $(\C + \R)$, and $m$ rounds of Amplitude Estimation $\Q = \A \S_0 \A^\dagger \S_{\psi_0}$. Read-out in all qubits is a requirement for post-selection to reduce errors. 

We will investigate in detail the performance of unary {\sl vs.} binary circuits for option pricing in Sec. \ref{sec:simulations}. There we will find out that the unary representation is advantageous to the binary one, when targeting the same realistic precision and errors are taken into account.

\section{Unary and binary comparison \label{sec:un-vs-bin}}

We compare here the unary algorithm for option pricing described in Sec. \ref{sec:unary} to the binary one stated in Sec. \ref{sec:binary}, in terms of the necessary circuit design as well as the number of gates required to apply the algorithm and successfully perform Amplitude Estimation.

\subsection{Ideal chip architecture}

\begin{figure}[t!]
    \centering
    \includegraphics[width=\linewidth]{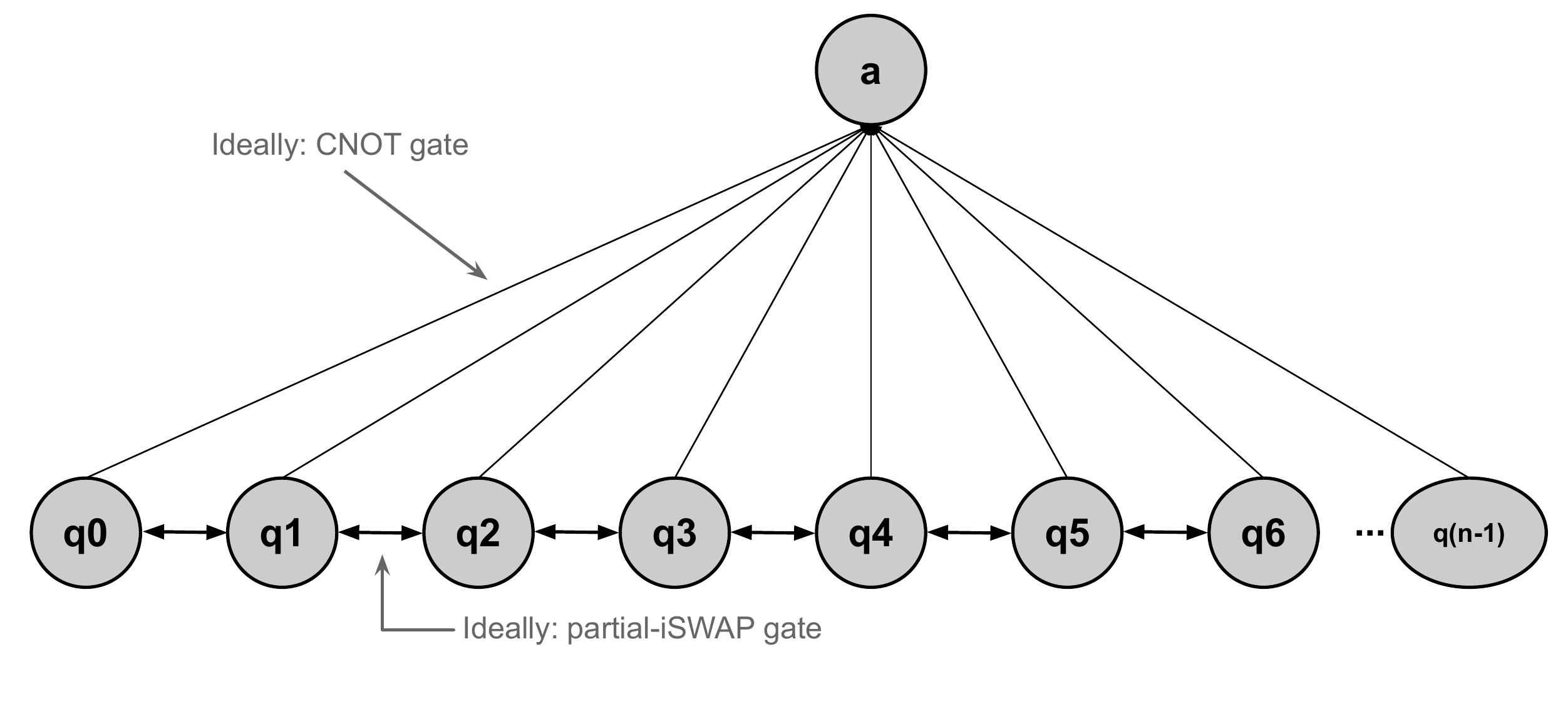}
\caption{Ideal chip architecture to implement the unary algorithm for option pricing. Only a single ancilla qubit, labelled as \emph{a} in the image, has to be non-locally controlled by the rest of the qubits. All other interactions are first-nearest-neighbor gates.}
\label{fig:ChipArchitecture}
      \includegraphics[width=.8\linewidth]{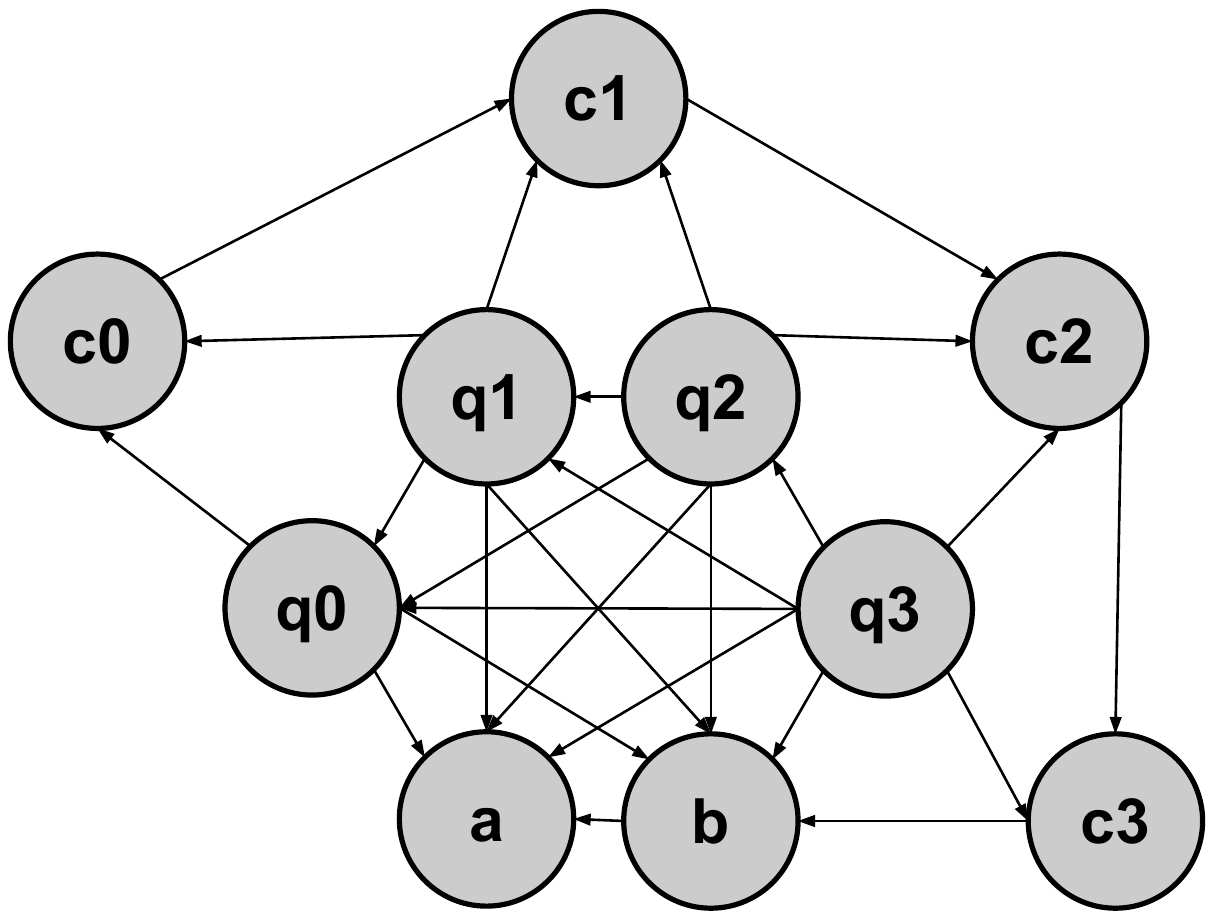}
\caption{Ideal chip architecture to implement the binary algorithm for option pricing with 4 qubits of precision, \emph{q}$_0$,\emph{q}$_1$,\emph{q}$_2$,\emph{q}$_3$, where \emph{a} and \emph{c} stand for ancillary and carrier qubit, respectively, and \emph{b} is another ancilla. The algorithm requires a number of ancillary and carrier qubits equal to the number of precision qubits plus two, 4+2 in this example. Full connectivity is needed between the precision qubits and two ancillas.}
\label{fig:ChipArchitectureBinary}
\end{figure}

The structure of the unary algorithm allows for a simple chip design. In order to upload the desired probability distribution to the quantum register, only local interactions between first-neighbor qubits are required. Therefore, qubits can be arranged on a single 1D line with two-local interactions. Such a connectivity is perfectly suited to carry out the algorithm. In order to compute the expected payoff, the ancillary qubit needs to interact with the rest of the quantum register. This structure is outlined in Fig. \ref{fig:ChipArchitecture} for an arbitrary number of qubits.

The simplicity of the architecture needed to implement the unary algorithm might yield an advantage over alternative algorithms in NISQ computers. Note also that superconducting qubits allow for a natural implementation of the partial-iSWAP gate \cite{partialiSWAP-bialczak2010}. This realization of the quantum circuit would result in a decrease in the number of needed gates by factor of 6 in the amplitude distributor module.

On the other hand, the binary algorithm for payoff calculation needs non-local chip connectivity. For the sake of comparison with the simplest chip architecture presented for the unary algorithm, the most basic connectivity needed to perform the steps described for the binary scheme is displayed in Fig. \ref{fig:ChipArchitectureBinary}.
The number of necessary qubits for the binary algorithm scales better asymptotically than in the unary approach, despite the increasing number of ancillary qubits required. Nevertheless, the need for Toffoli gates and almost full connectivity may eliminate this advantage in practical problems for NISQ devices.

\subsection{Gate count\label{subsec:counting}}

\begin{table*}[t]
    \centering
    \resizebox{0.38\textwidth}{!}{
    \begin{tabular}{|l|c|c|c|c|c|c|c|c|}\hline
        \multirow{2}{*}{\bf Unary} & \multicolumn{4}{c|}{CNOT} & \multicolumn{4}{c|}{partial-iSWAP} \\ \cline{2-9}
         & $\D$ & $\C + \R$ & $\S_{\psi_0}$ & $\S_0$ &$\D$ & $\C + \R$ & $\S_{\psi_0}$ & $\S_0$\\ \hline
        1-qubit gates & 2n & 2$\kappa$n & 1 & 4 & 1 & $\kappa$10n & 1 & 9\\
        2-qubit gates & 4n & 2$\kappa$n & 0 & 1 & n & $\kappa$5n & 0 & 2\\\hline
        Circuit depth & 3n & 4$\kappa$n & 1 & 5 & n/2 & 15$\kappa$n & 1 & 10 \\\hline
    \end{tabular}} 
    \hfill
    \resizebox{0.55\textwidth}{!}{
    \begin{tabular}{|l|c|c|c|c|c|c|c|c|}\hline
        \multirow{2}{*}{\bf Binary} & \multicolumn{4}{c|}{CNOT} & \multicolumn{4}{c|}{partial-iSWAP} \\ \cline{2-9}
         & $\D$ & $\C + \R$ & $\S_{\psi_0}$ & $\S_0$ & $\D$ & $\C + \R$ & $\S_{\psi_0}$ & $\S_0$ \\\hline
        1 qubit gates & 3nl  & (16+5$\kappa$)n & 1 & 20n - 23 & 8nl & (86+5$\kappa$)n & 1 & 80n - 113 \\
        2 qubit gates & nl & 14n & 0 & 12n - 18 & 2nl & 28n & 0 & 24n - 36\\\hline
        Circuit depth & nl+l & (27+2$\kappa$)n & 1 & 24n - 30 & 6nl+l & (97+2$\kappa$)n & 1 & 90n - 129\\\hline
    \end{tabular}}
    \caption{Scaling of the number of 1- and 2-qubit gates and circuit depth as a function of the number of qubits $n$ representing the asset value in unary and binary representations, for the amplitude distributor $\D$, payoff estimator $\C + \R$ and Amplitude Estimation operators $\S_{\psi_0}$ and $\S_0$. Ideal chips architectures are assumed. We compare this scaling in case CNOT or partial-iSWAP gates are implemented. In case the experimental device can implement both CNOT and partial-iSWAP basic gates, the total amount of gates and total depth would be reduced. For the unary circuit, the parameter $0\le \kappa\le 1$ depends on the position of the strike in the qubit register. In the binary case, note the large overheads due to the use of Toffoli gates. The parameter $0\le \kappa\le 1$ characterizes the number of $1$s in the binary representation of the strike price. For the amplitude distributor, $l$ is the number of layers of the qGAN. 
    }
    \label{tab:gates}
\end{table*}

\begin{figure*}[t!]
    \centering
    {\includegraphics[width=0.325\textwidth]{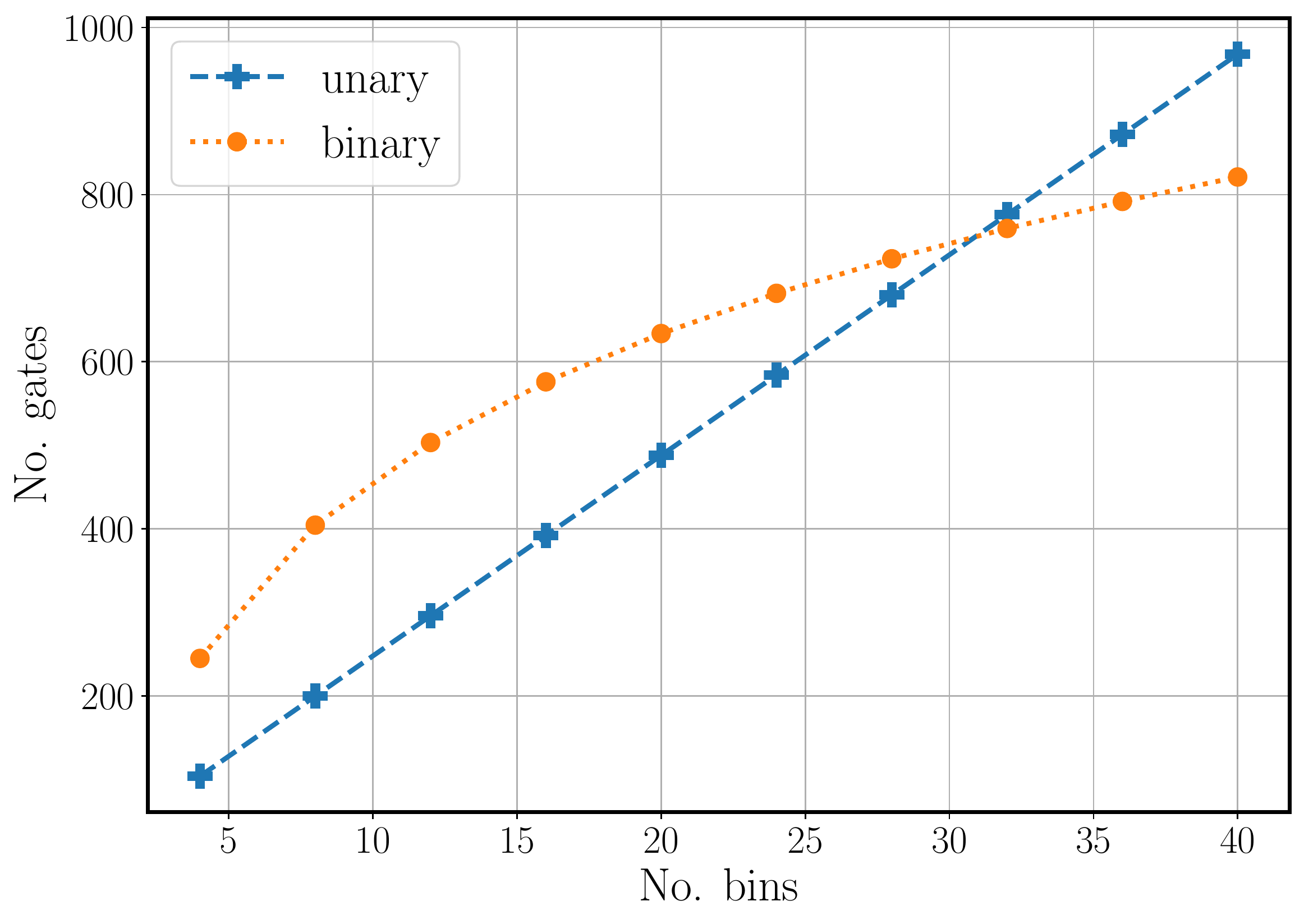}}
    {\includegraphics[width=0.323\textwidth]{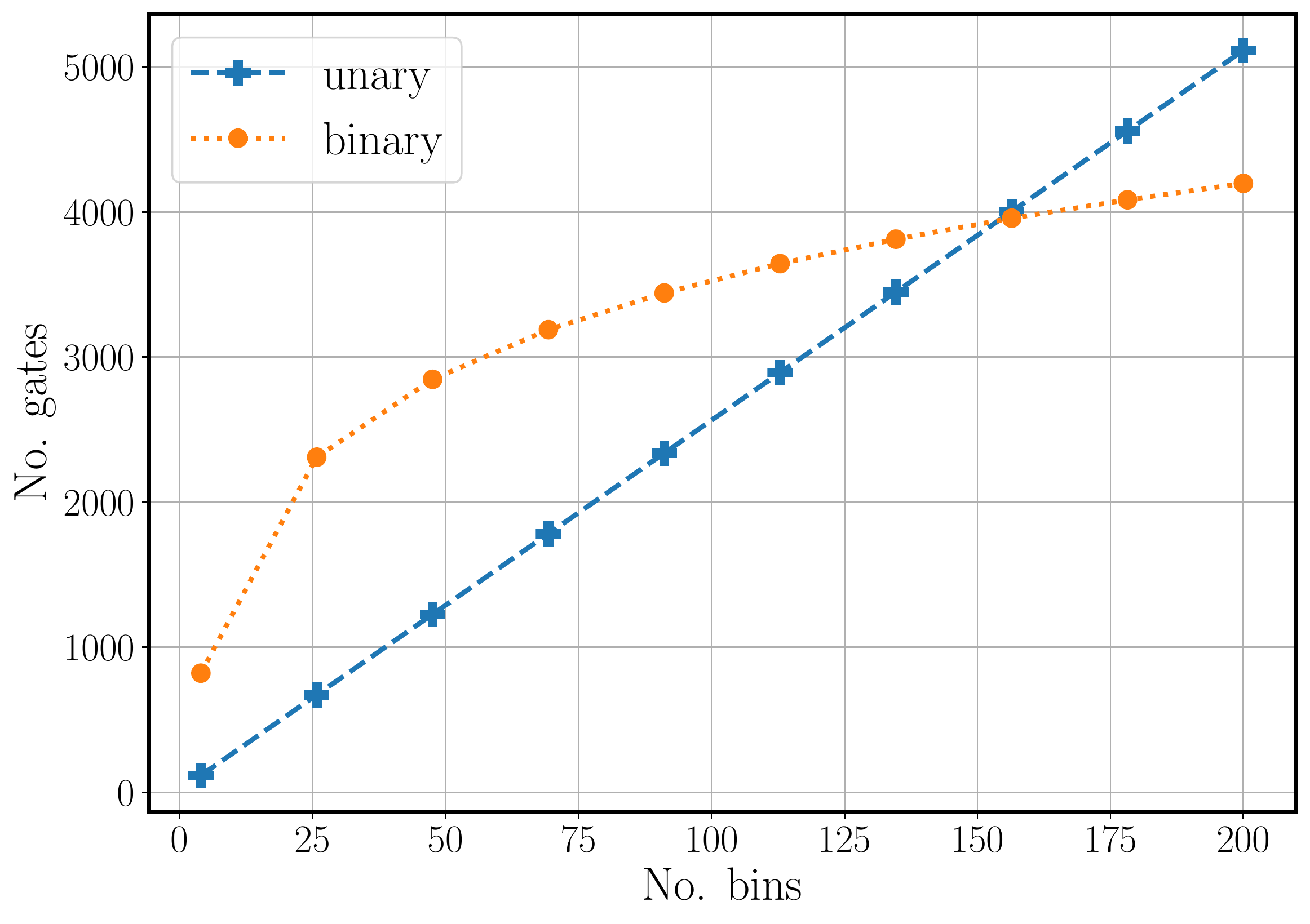}}
    {\includegraphics[width=0.317\textwidth]{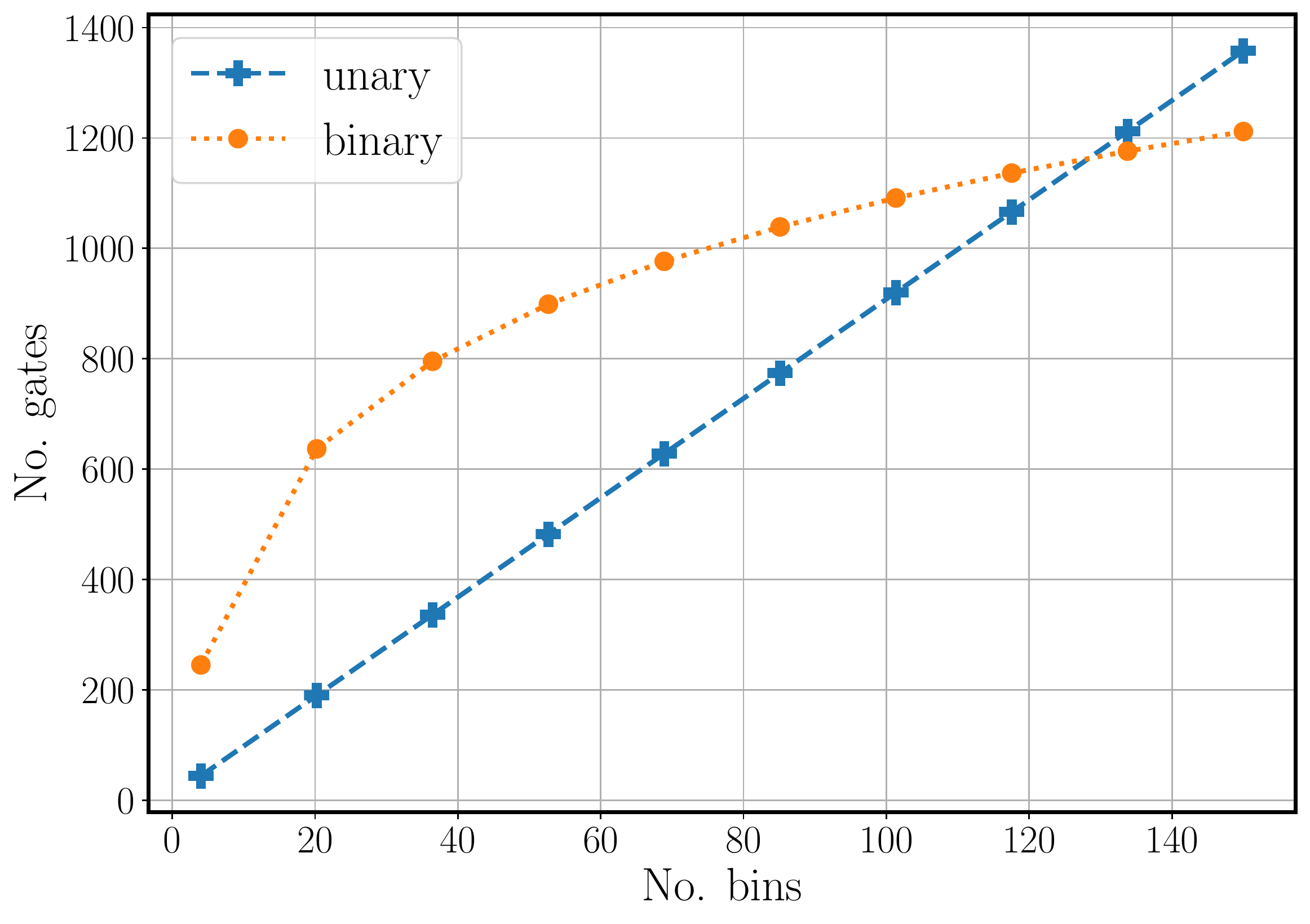}}
    \caption{Scaling of the number of gates required for the full algorithm, including a step, $m=1$, of Amplitude Estimation, with the number of bins, for different native gates: CNOT gates (Left), partial-iSWAP gates (Center) and the best possible combination (Right), in which one is allowed both CNOT and iSWAP gates as native to the device. The scaling is calculated assuming ideal connectivity, which would largely hinder the binary implementation were that not the case.}
    \label{fig:gate_count}
\end{figure*}

The unary algorithm needs $\order{n}$ partial-SWAP gates in order to distribute the amplitude and $\order{\kappa n}$ controlled-$R_y$ gates to encode the payoff in an ancillary qubit, where $0\le \kappa\le 1$ depends on the strike price K. However, actual quantum devices operate using a native set of gates that are used to construct any other unitary. We present in Table \ref{tab:gates}, left, the gate count of the full circuit as a function of the number of qubits, using either CNOT or partial-iSWAP as the native entangling gate. The gate count assumes the simple ideal chip structure, see Fig. \ref{fig:ChipArchitecture}, that requires first-neighbor interactions and an ancilla connected to the rest of the qubits.

The partial-iSWAP gate introduces a substantial gain for the amplitude distributor but requires more gates in order to implement the payoff calculation. If both partial-iSWAP interaction between nearest neighbors and CNOT-based connection with the single ancilla are implemented, the best scaling of the full algorithm would be achieved. To be precise, the total number of gates would be $(4 \kappa+1) n+1$, and the depth of the circuit would become $(4 \kappa +\frac{1}{2})n$.

The scaling of the gate count for the binary algorithm is displayed in Table \ref{tab:gates}, right. We compare the scaling when using CNOT or partial-iSWAP as native gates. The  CNOT gate turns out to be more convenient for the binary algorithm. These results include the part of the algorithm that produces the uploading of the probability distribution into the quantum register (hence the dependence on the number of layers of the variational circuit), but it does not take into account the training required by the qGAN. In both unary and binary cases, the counting for single-qubit gates was made compiling several successive single-qubit gates into a single one.

This gate count is performed assuming full connectivity, or at least the connectivity presented in Fig. \ref{fig:ChipArchitectureBinary}. Existing quantum devices need to implement extra SWAP gates to account for insufficient connections, which are not taken into account in this calculations. Therefore, the gate counting on a computer with less than this ideal connectivity will result in a worse scaling.

Let us emphasize that the gate overhead for the unary representation is much lower than the one for the binary case. This is due to the fact that the unary circuit does not require any three-qubit gate. This simplification is eclipsed by the gain in precision for large $n$, provided an efficient uploading of probability distributions is found for the binary case. The detailed gate count comparing unary {\it vs.} binary circuits is shown in Fig. \ref{fig:gate_count}, where we have taken $\kappa=\frac{1}{2}$ and $l=\frac{\log_2 n}{2}$, where $l$ is the number of layers of the qGAN. In order to compare like with like, the comparison of scaling is made as follows. For a given number of $n$ bins, which directly relate to precision, we take $n$ qubits in the unary representation and only $\log_2 n$ in the binary one. Note that the overhead in the binary representation makes the unary one more convenient for a number of bins less than $\sim 100$. This scaling behaviour confirms that this unary representation would get outperformed by the binary one for a large number of bins, provided the devices performed gates with no error. However, if quantum resources are limited, as in NISQ devices, circumstances are favorable for the unary representation. Moreover, in practise, the connectivity requirements further benefits the unary representation over the binary one. 

\section{Simulations\label{sec:simulations}}

The circuits we present in this paper can be simulated using the tools provided by the Python package \textit{Qiskit} \cite{Qiskit}. We first consider the unary and binary algorithms in ideal conditions, that is, we verify the performance of the quantum circuits in the absence of any noise. Then, we test them both under increasing amounts of different sources of noise in order to assess which of the two procedures may be more advantageous for NISQ devices. 

The simulations in this work were carried out using a simple yet descriptive model. In the case of single-qubit and two-qubit gate errors, we consider depolarizing noise. That is equivalent to transforming the state after each gate application by $\rho \rightarrow (1 - \epsilon) \rho + \epsilon Tr \rho \frac{I}{d}$, with $d$ the dimension of the Hilbert space. Measurement errors are ten times more likely to happen $(10\epsilon)$, and they are symmetric, \ie the probability of measuring an incorrect $\ket 0$ or $\ket 1$ is identical. Let us remark here that we have not included thermal relaxation or thermal dephasing. The reason is that, given the shallow depth of the simulated circuits, the execution times are far below current coherence times of qubits (the latter being $\sim 1000$ times the duration of a single-qubit gate), and thermal errors are therefore negligible. This description was adjusted to be comparable to state-of-the-art quantum devices \cite{supremacy2019}.

The accuracy of the expected payoff estimation is used to benchmark both algorithms against the aforementioned errors, as a function of the interpolation parameter $\epsilon$. The simulations were performed with 8 and 3 qubits in the unary and binary basis, respectively, using their ideal chip structures, see Sec. \ref{sec:un-vs-bin}. Notice that both cases correspond to 8 bins. Recall as well that the unary approach includes post-selection, which results in a clear improvement of the algorithm's performance.

The results presented in this section consider depolarizing and measurement errors together. A separate analysis of these errors can be found in App. \ref{sec:ap_more_results}. The code is publicly available in Ref. \cite{github}. It allows to perform simulations with different combinations of several errors, namely {\sl bitflip}, {\sl phaseflip}, {\sl bitphaseflip}, thermal and measurement errors, isolated or as part of  custom error models.

\subsection{Amplitude distribution loading}

The log-normal probability distribution used for the simulations is generated in accordance with the Black-Scholes model discussed in Sec. \ref{sec:econ_model}. We work with a particular example, chosen such that the asset price at $T=0$ is $S_0=2$, the volatility of the asset is $\sigma=40\%$, the risk-free market rate is $r=5\%$, the maturity time is $T=0.1$ years, and the accorded strike price for the asset is $K=1.9$. The simulation of the asset price ranges up to three standard deviations from the mean value of the distribution.

\begin{figure}
    \centering
    \includegraphics[width=\linewidth]{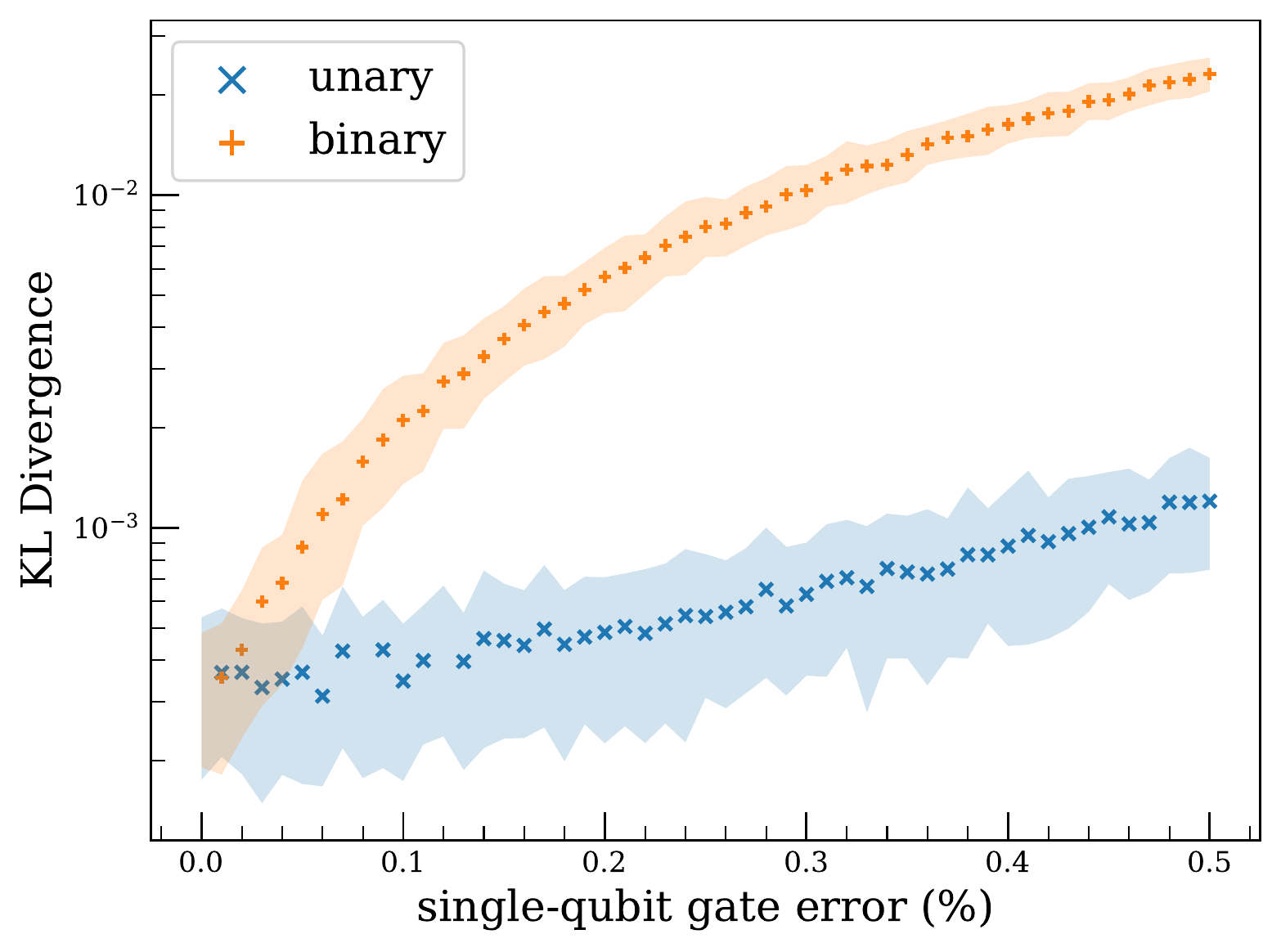}
    \caption{Kullback-Leibler divergence between the target probability distribution and those achieved by the quantum algorithms, for equivalent 8 unary qubits and 3 binary qubits, and different levels of depolarizing error. Crosses stand for average results, and the shaded regions encompass the central $70\%$ of the instances. Each probability distribution is estimated using $100$ experiments with $10^4$ samples each. For noiseless computers, the KL divergence almost vanishes, but gets larger as noise is added. For the maximum allowed error, the KL divergence of the binary algorithm is one order of magnitude larger than that of the the unary one.}
    \label{fig:KL}
\end{figure}

The capability of quantum computers to approximate a given probability distribution in the presence of noise can be quantified by the Kullback-Leibler (KL) divergence \cite{KL-kullback1951}. This quantity measures the distance between two probability distributions, vanishing when they are indistinguishable. Fig. \ref{fig:KL} plots the KL divergence for the unary and binary approximations to the log-normal distribution. For the maximum allowed error, the KL divergence of the binary algorithm is one order of magnitude larger than that of the unary one.

\subsection{Expected payoff calculation}

In terms of payoff calculation, the algorithms diverge slightly. Classically, with a precision of $10^4$ bins, the estimated payoff for this financial option is $0.1595$, that we take as the exact value for comparison with the quantum strategies. Recall that in order to compare like with like, on the quantum side we have 8 unary qubits and 3 binary qubits, that both correspond to 8 bins. 

\begin{figure}[t]
    \centering
     \label{fig:bin_error_classical}{\includegraphics[width=.9\linewidth]{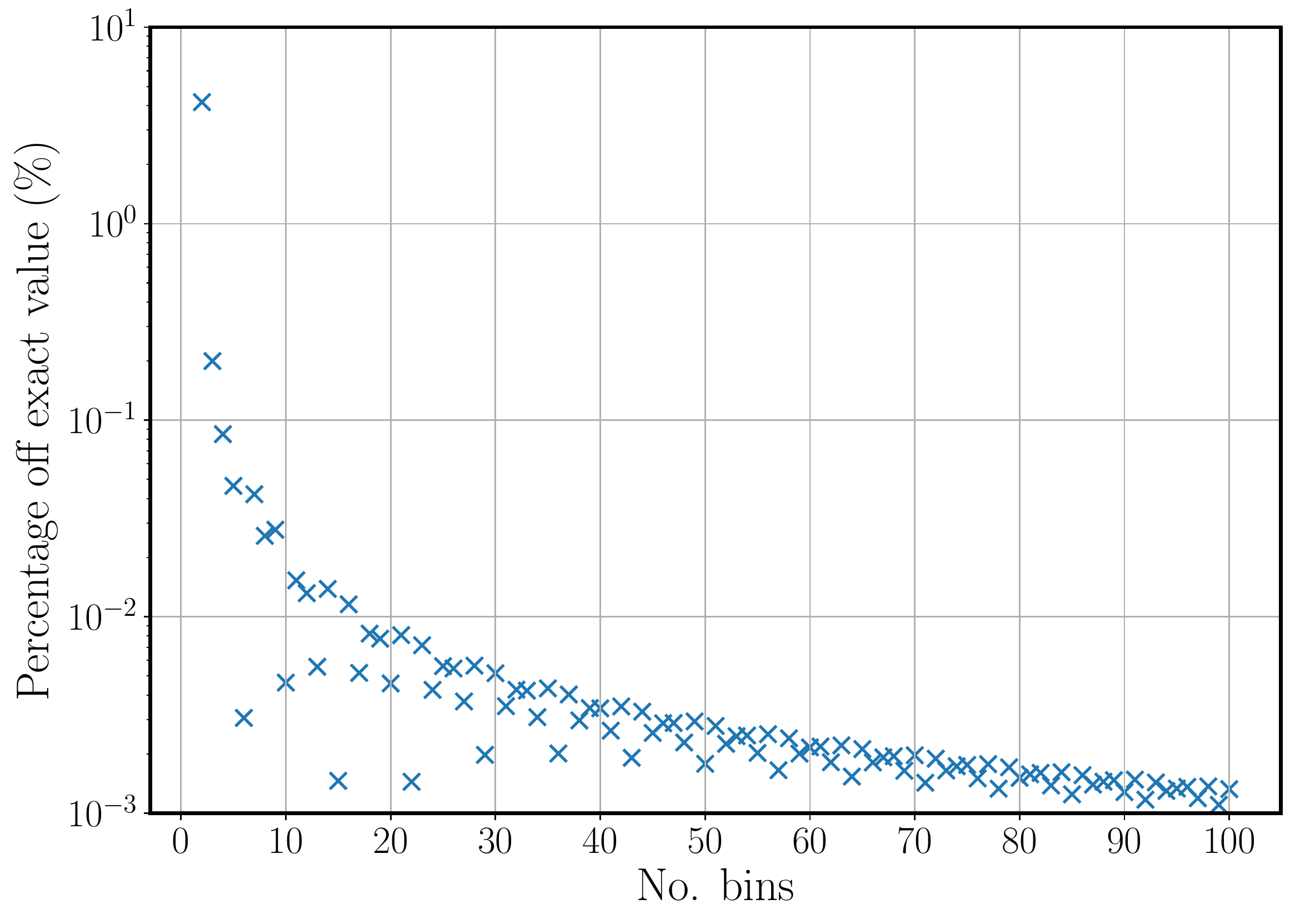}}
    \caption{Percentage error from the exact value of the expected payoff, for the classical computation, as a function of the number of bins in the probability distribution.  With only $\sim 50$ bins, errors for the option price below 0.5\% are already reached.}
    \label{fig:bin_error}
\end{figure}

In Fig. \ref{fig:bin_error}, we show the error of the expected payoff as a function of the number of bins in the probability distribution, for the classical computation. This precision depends on the binning and the position of the strike. Therefore, at a large enough number of bins, the results fall within a reasonable percentage of the actual value. At 100 bins, errors for the option price go well below 1$\%$. This shows that the unary algorithm can be implemented in the range where it uses less quantum gates than the binary algorithm, and still have low discretization errors coming from the binning.

\subsubsection*{Robustness against noise}

\begin{figure}[t]
    \centering \includegraphics[width=0.9\linewidth]{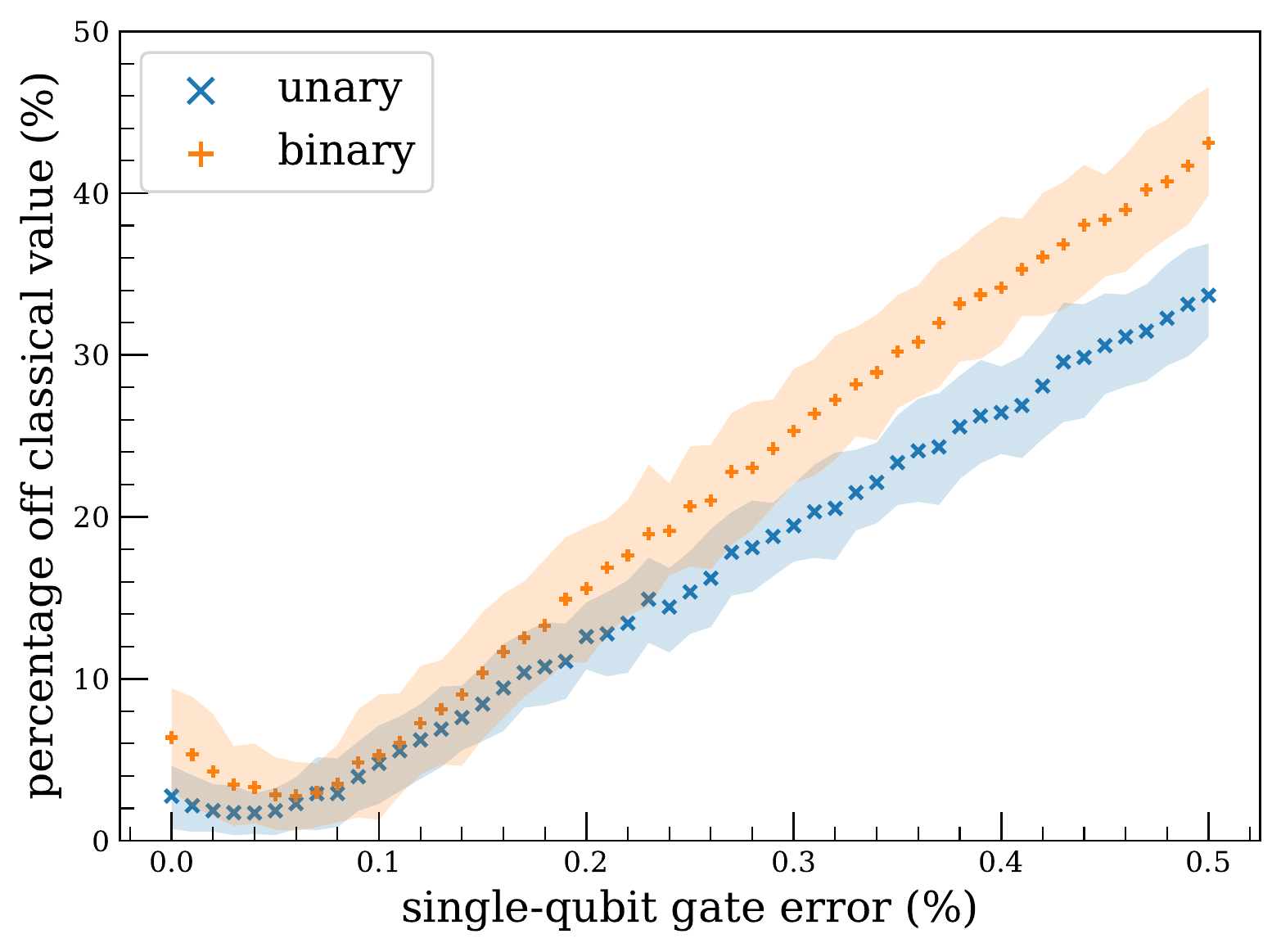}
    \caption{For equivalent 8-qubit unary and 3-qubit binary algorithms, percentage error in the payoff calculation for depolarizing and measurement errors, up to 0.5\% for single-qubit gates, 1\% for two qubit gates and 5\% for read-out errors, consistent with state-of-the-art devices. The calculations were averaged over $100$ repetitions with $10^4$ runs each. The shaded regions encompass the central 70\% of the instances in each case. The unary algorithm is more robust against these errors.
    }
    \label{fig:depolarizing_m_error}
\end{figure}

We show in Fig. \ref{fig:depolarizing_m_error} the average of the relative error of the expected payoff computation when compared to the classical value. The $x$ axis of Fig. \ref{fig:depolarizing_m_error} depicts the single-qubit gate error percentage $\epsilon$, but two-qubit and read-out errors are also included following the model explained previously.
The shaded region includes $70\%$ of the total instances used for the average. It can be seen that the unary algorithm, in general, has less deviation form the mean value than the binary algorithm.

\begin{figure*}
    \centering
    \includegraphics[width=0.5\textwidth]{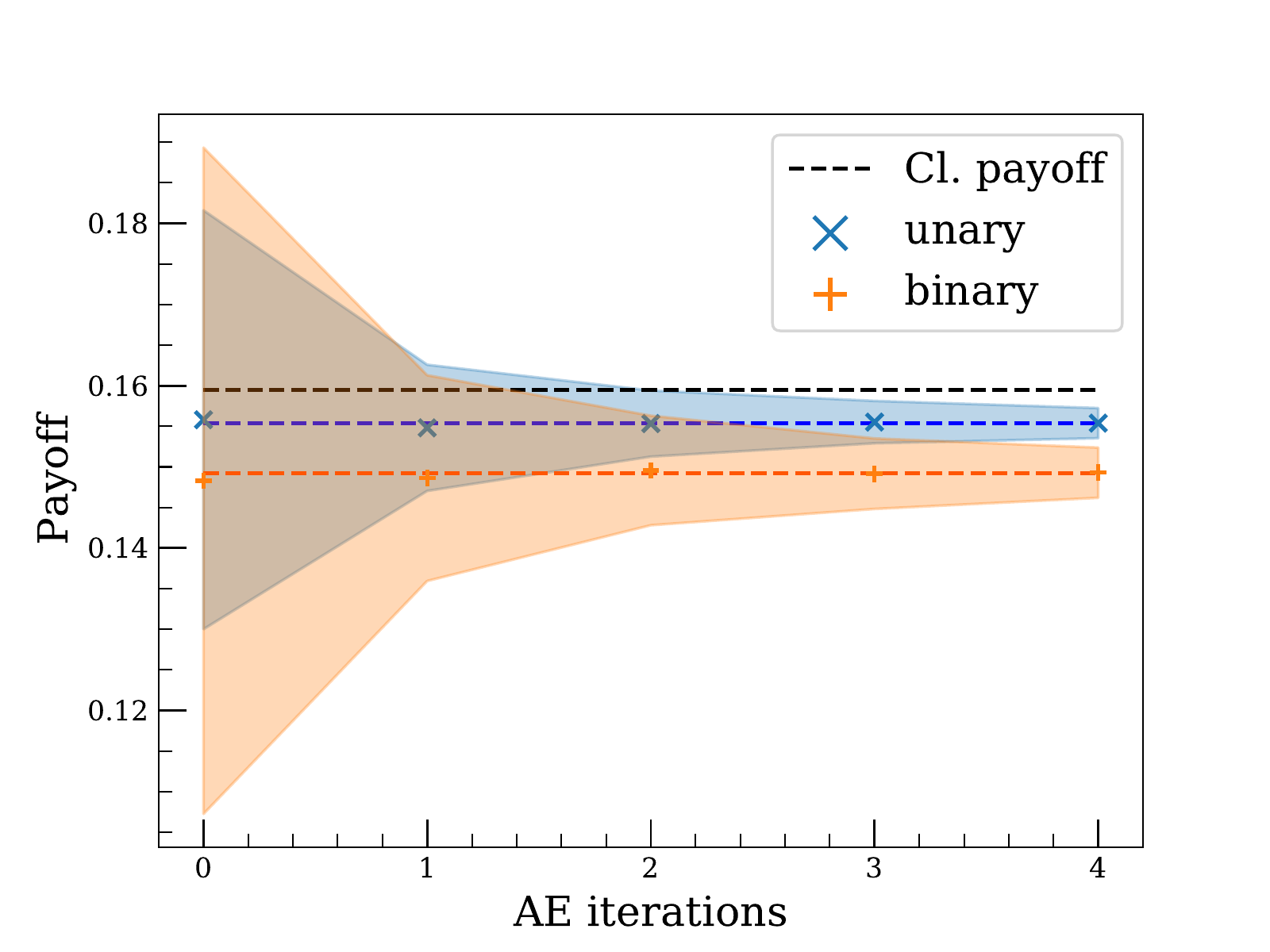}
    \hfill {\includegraphics[width=0.45\textwidth]{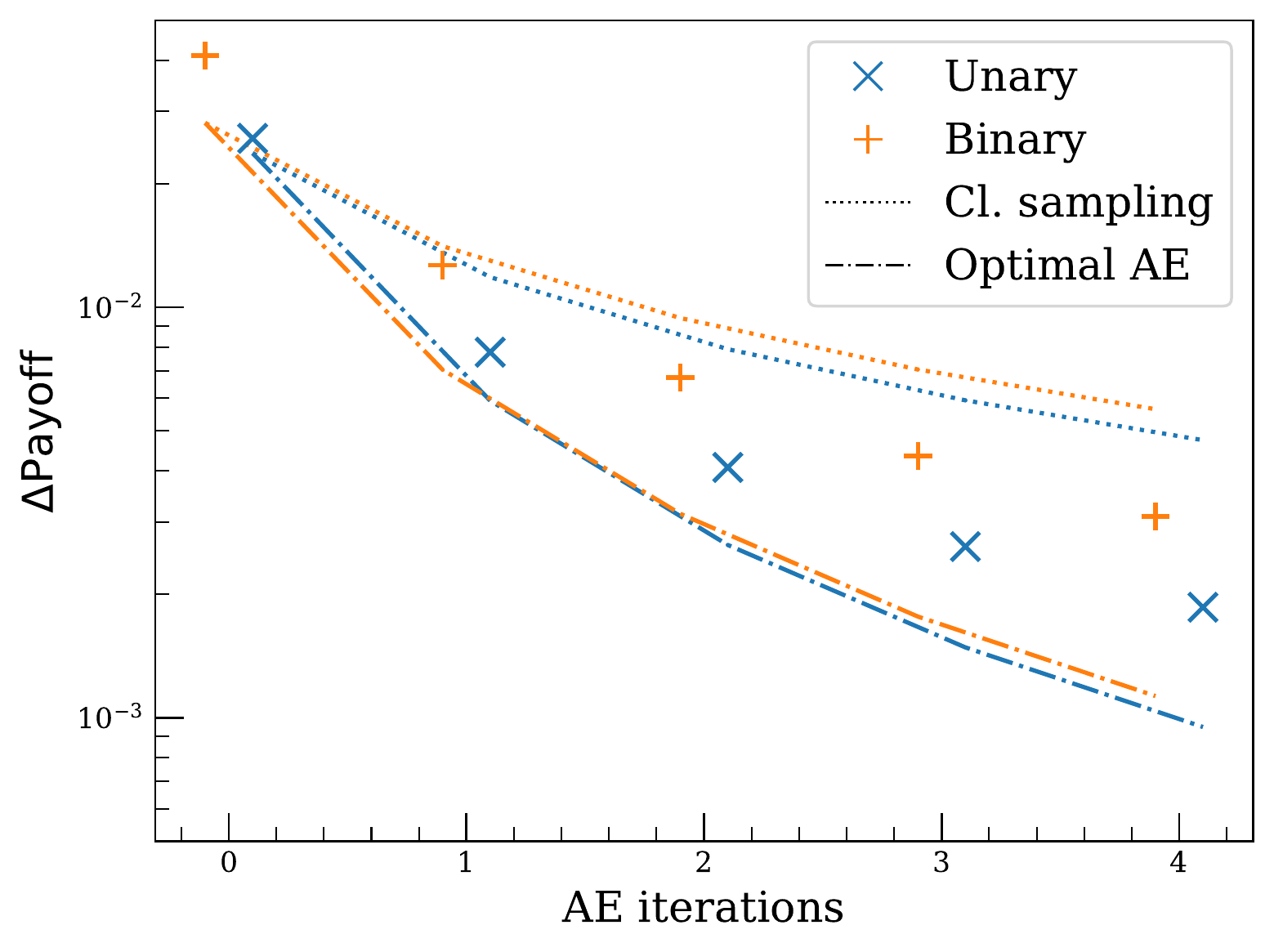}}
    \caption{Left: Mean and uncertainty of the outcomes of the expected payoff, obtained with Eq. \eqref{eq:ap_exp_errors} and proper transformations. The dashed lines indicate the exact values. Unary and binary approaches are depicted, and convergence to the optimal values are obtained for both. Notice that these values are not the same since the outcomes of both algorithms are not equally related to the payoff. The shaded regions correspond to the statistical uncertainty. Right: Statistical uncertainties in the expected payoff. The dotted lines indicate the uncertainty given by classical sampling, while the dot-dashed lines represent the optimal uncertainty provided by Amplitude Estimation. Results of the simulations lie in between. In this figure we compare procedures with the same number of applications of the $\A$ or $\A^\dagger$ operators, for noiseless circuits.}
    \label{fig:convergence_results}
\end{figure*}

\subsection{Amplitude Estimation}

This section comprises results obtained for the Amplitude Estimation algorithm, that can be divided in three parts. First, results are shown for the case of noiseless devices, converging to the expected value within errors due to binning and Taylor approximation, the latter only in the binary case. All results were obtained for 8 bins, unless stated otherwise. Second, an analysis on the effect that quantum errors induce on the estimated value of $a$ has been performed, both for unary and binary approaches. Third, an analysis on the statistical uncertainty incurred in the estimation has been also included. 

Only Amplitude Estimation without phase estimation can be performed on NISQ devices. In these simulations, we have used a procedure based on weighted averages that consider both mean values and uncertainties, for a given series of AE steps, see App. \ref{sec:ap_iae} for further details. In our results, every instance has been repeated $100$ times. The choice of $m_j$ is linear, $m_j = j$, with $j=\{0, 1, 2, \ldots\}$, in order to control how the performance evolves. The confidence level was adjusted to $1 - \alpha = 0.95$.

\begin{figure*}[t!]
    \centering
    {\includegraphics[width=.45\textwidth]{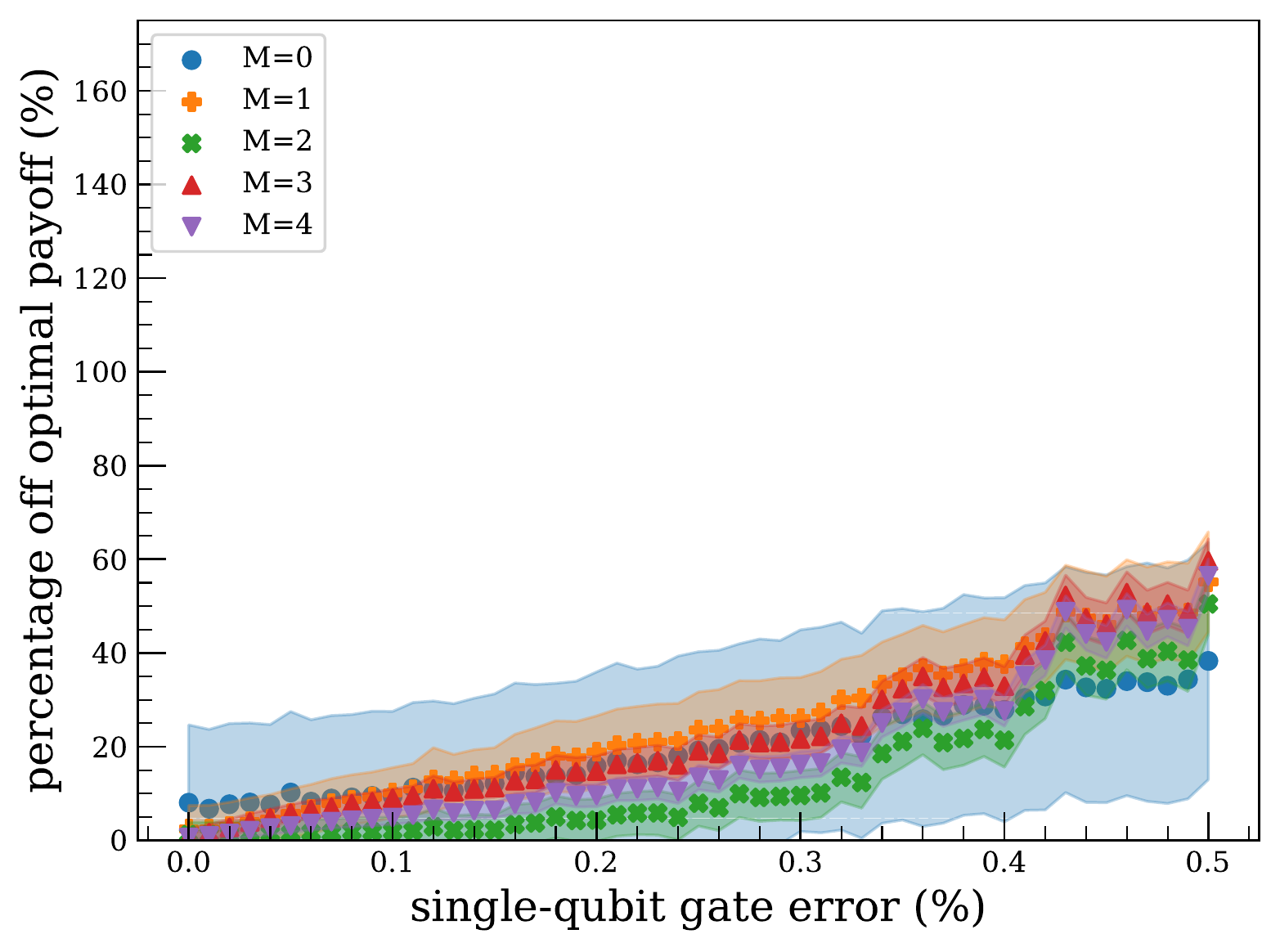}}
    \hfill {\includegraphics[width=.45\textwidth]{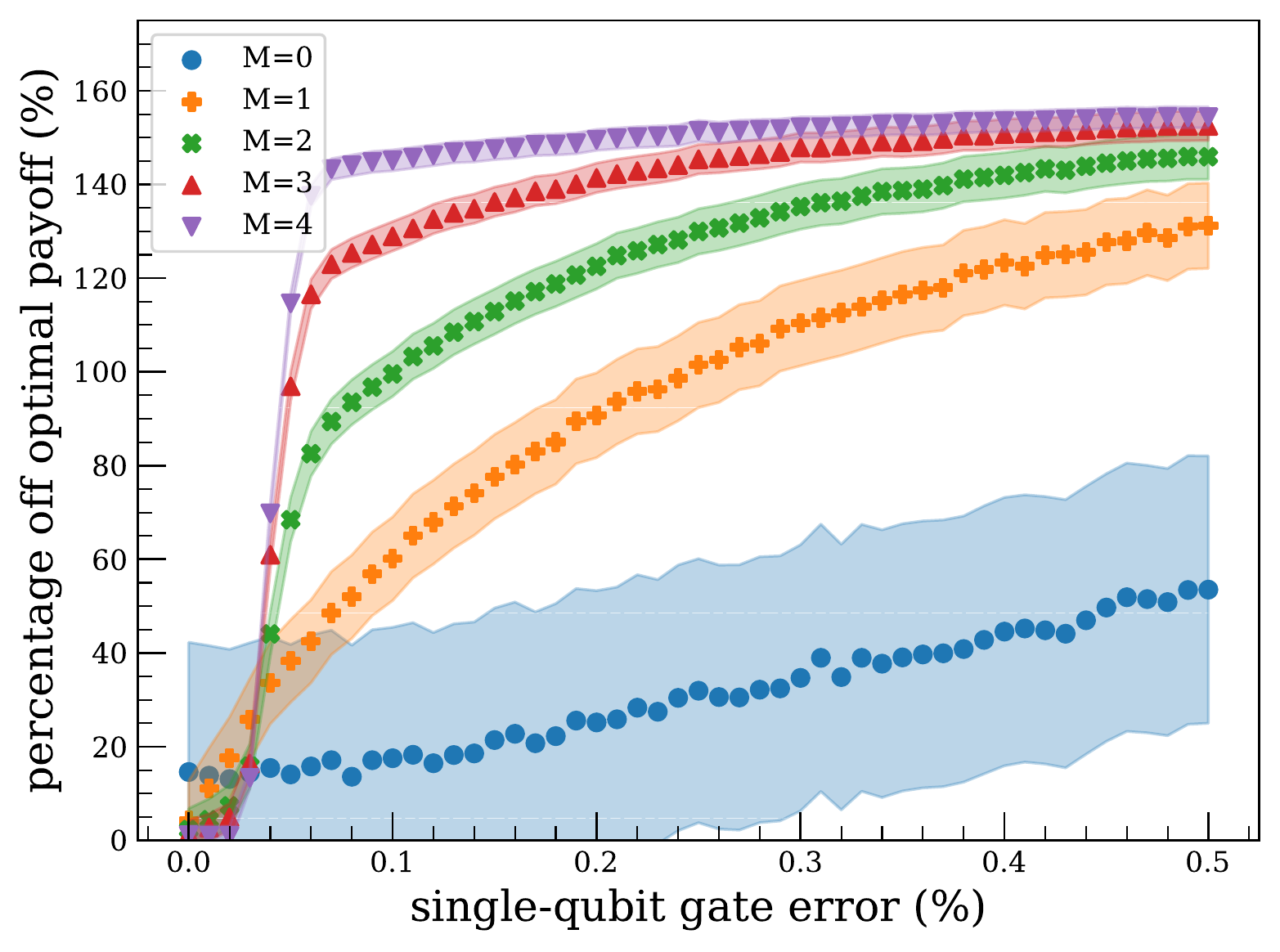}}
    \caption{Results of the errors in the expected payoff respect to the optimal value, for the unary (Left) and binary (Right) representation, with $M$ iterations of Amplitude Estimation for both the unary and binary approaches. Depolarizing and read-out errors have been considered. Scattering points stand for average values, while the shaded region corresponds to the statistical uncertainties of the results. The behavior of both approaches is very different. In the unary case, the expected payoff is resilient to errors. On the other hand, the binary approach returns acceptable results for $M=0$, while $M\geq 1$ rapidly saturates to the outcome of a random circuit.}
    \label{fig:data}
\end{figure*}

In Fig. \ref{fig:convergence_results} it is shown how Amplitude Estimation increases the precision of the measured outcome, converging to the actual value as more iterations of AE are used. The results of this simulation unveil that Amplitude Estimation reduces with every iteration the uncertainty in the value of the expected payoff. 

The next step of the analysis is to assess the robustness of both the unary and binary representation against noisy circuits. The results for the deviations in the outcomes of $a$ obtained for noisy circuits are depicted in Fig. \ref{fig:data}, taking into account depolarizing and read-out errors together. The number of iterations has been limited to $M=4$. Two very different behaviors can be observed. In the case of the unary approach, the outcomes endure the noise of the device for $M=0,1,2,3,4$ and for low error rates, while entering into an erratic regime for large ones. For instance, at $M=2$, a result that is very close to the optimal value and with low uncertainty is obtained up to error parameter $\epsilon \sim 0.3\%$. In contradistinction, the binary approach loses its robustness for very small noise levels and $M \geq 1$. This can be attributed to the post-selection scheme and to the lower number of applied gates, that benefit the unary algorithm significantly. The decrease of the uncertainties is detailed in Fig. \ref{fig:errors} in the Appendix.

From these results we can infer that the Amplitude Estimation procedure, when performed on NISQ devices, provides a quantum advantage only for the unary representation and for limited noise levels in the device; specifically, similar to those present in available state-of-the-art machines \cite{supremacy2019}.

\begin{figure}[t!]
\centering
    \includegraphics[width=\linewidth]{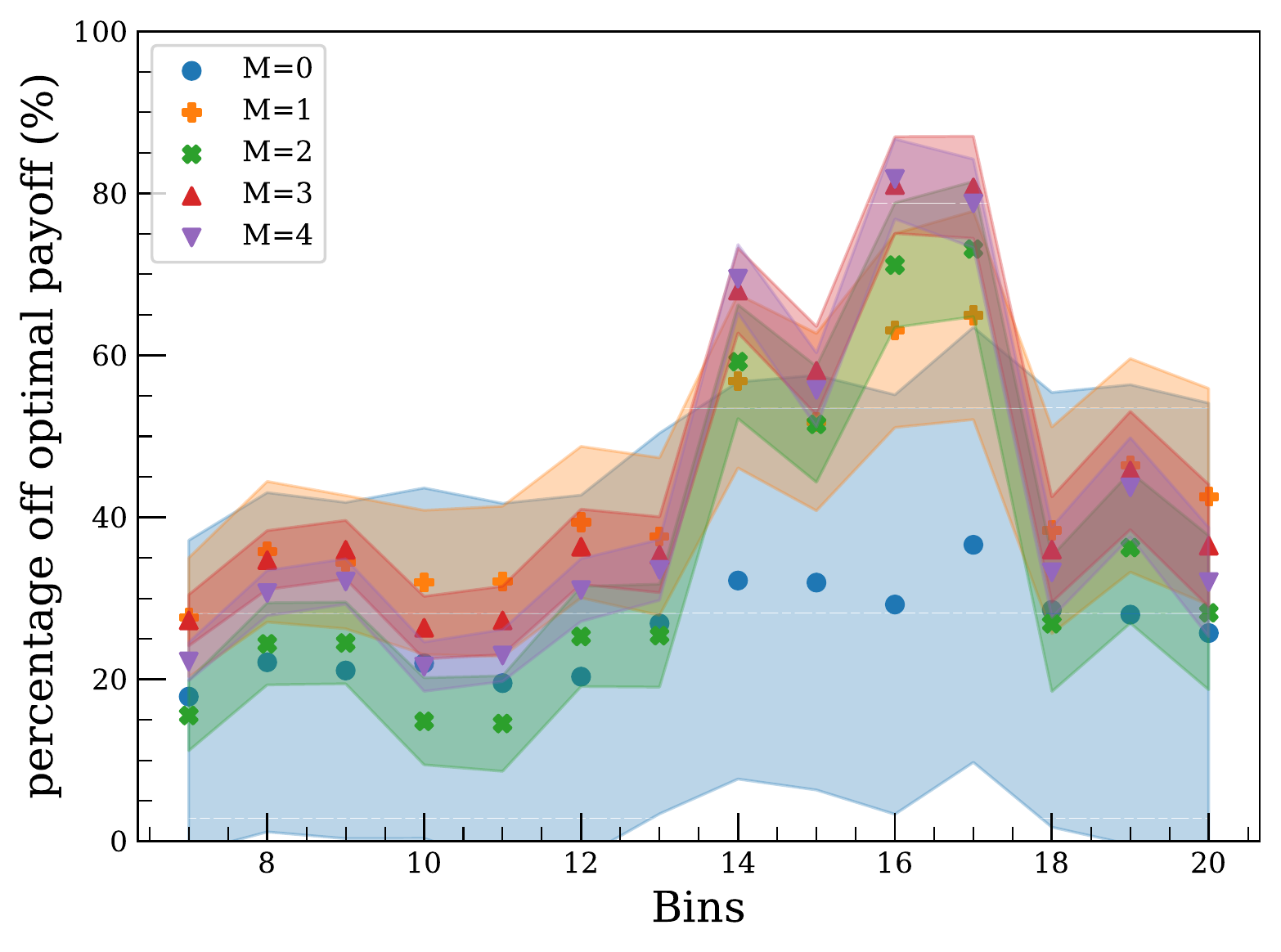}
    \caption{Results for the error of the expected payoff for increasing number of bins for up to $M$ iterations of Amplitude Estimation for the unary approach, considering depolarizing and read-out errors together. Scattering points represent the mean values obtained for the experiment, while shadowed areas include 70\% of the instances. } 
    \label{fig:several_bins}
\end{figure}

Simulations have been extended to several different numbers of bins in the unary representation. We show in Fig. \ref{fig:several_bins} how the deviation in the payoff from the exact value evolves when larger quantum systems are taken into account. In this example, the error parameter was adjusted to $\epsilon = 0.3\%$. We have considered depolarizing and measurement errors. Each experiment was repeated only 10 times to reduce computational costs. In Fig. \ref{fig:several_bins} it is possible to see that the deviation in the payoff increases as more bins are added. This corresponds to the expected trend since systems with more qubits require a larger number of gates, and thus errors are more likely to happen. Larger errors are observed for numbers of bins between 13 and 18. This behavior is expected to reach a saturation regime for large enough error rates. In the binary case, since the circuit is prone to large errors, the output becomes indistinguishable from one of a random circuit. However, this regime has not been reached yet in the unary representation. Concerning the sampling uncertainty, it is larger as the number of bins increase. This reflects that the more bins, the more errors may occur, and thus more instances are to be discarded via the post-selection mechanism, which translates into a slower convergence rate. The decrease of the uncertainties is detailed in Fig. \ref{fig:several_bins_2} in the Appendix.

\section{Conclusions} \label{sec:conclusions}

Finance stands as one of the fields where quantum computation may be of relevance. We have here presented a quantum algorithm that allows for the pricing of European options whose defining trait is that it works in the unary representation of the asset value.

We have illustrated our algorithm in the particular case of a single European option, whose maturity price for the underlying asset is obtained as the solution of a Black-Scholes equation and its expected return depends on a prefixed strike value. The global structure of our algorithm is divided in three steps: a) generation of the amplitude distribution of the asset value at maturity, b) evaluation of the expected return given the strike value, and c) Amplitude Estimation. Our algorithm relies on several new ideas to make this strategy concrete. 

The very first step is to define the level of precision the algorithms should aim at. This precision is related to $n$, the number of qubits in the circuit. The more qubits, the more resolution we can get. 

The next step corresponds to building the amplitude distribution of the asset at maturity. We have proposed to handle this problem using a circuit of depth $n/2$ that operates as a distributor of probability amplitude. Given a classical description of the probability distribution, this step of the algorithm substitutes the classical Monte Carlo generation of probabilities.

The computation of the expected return is particularly simple in the unary representation. It only needs a series of $n$ conditional two-body gates from the original qubit register to an ancilla. It then follows iterative Amplitude Estimation.

The use of unary representation seems at odds with performing precise computations. This is not so, as the precision of the expected return is an average over a sampled probability distribution which does not need to have too high a resolution. We verify this statement in detail to find that less than 100 qubits are enough to have competitive computations. 

The unary algorithm aims to be used during a middle stage between current quantum computers and fault-tolerant devices. The algorithm is designed to find applicability with a relatively low, while still useful, number of bins. Thus, we have designed a circuit which is simple in terms of logic operations and requires a much less sophisticated connectivity than its binary counterpart.

Unary representation definitely offers relevant advantages over the binary one. First, it allows for a simple distribution of probability amplitudes. Second, it provides a trivial computation of expected returns. Third, unary representation should only trigger one output qubit, while reading the expected return in the ancilla. This offers a consistency check. If no output, or more than one are triggered, the run is rejected. The ability to post-select faithful runs mitigates errors and increases the performance of the quantum algorithm. In addition, Amplitude Estimation may be performed successfully only in the unary basis, considering error levels in NISQ devices, since the procedure is more resilient to errors than the binary one. 

There are a number of further improvements that may be included in the algorithm. It is possible, for instance, to increase precision by taking the qubits to represent non equispaced elements in the probability distribution. It is enough to populate more densely the subtle regions of the sample distribution to gain some precision. Ideas to include multi-asset computations are also available \cite{underconstruction}. 

Finally, let us mention that our unary option pricing algorithm could be tested experimentally on quantum computers recently presented.

\section*{Acknowledgements}

SRC, APS, DGM, CBP and JIL are supported by Projects PGC2018-095862-B-C22 and Quantum CAT (001-P-001644). APS, DGM, CBP and JIL acknowledge CaixaBank for its support of this work through Barcelona Supercomputing Center's project CaixaBank Computación Cuántica. 

\section*{Code availability}
The code is available in \href{https://github.com/UB-Quantic/quantum-finance}{Github} \cite{github}.

\bibliographystyle{apsrev4-1}
\bibliography{Citations}

\newpage

\onecolumngrid

\appendix
\section{Classical Option Pricing\label{sec:ap_econ_model}}

The evolution of asset prices in financial markets is usually computed using a model established by F. Black and M. Scholes in Ref. \cite{blackscholes-black1973}. This evolution is governed by two properties of the market, the interest rate and the volatility, which are incorporated into a stochastic differential equation. The equations controlling a set of assets are usually solved using Monte Carlo methods. 

\subsection{The Black-Scholes model}\label{subsec:Black-scholes}

The Black-Scholes model for the evolution of an asset is based on the following stochastic differential equation \cite{blackscholes-black1973}
\begin{equation}\label{eq:ap_BSM}
    {\rm d}S_T = S_T\, r\, {\rm d}T + S_T\, \sigma\, {\rm d}W_T, 
\end{equation}
where $r$ is the interest rate, $\sigma$ is the volatility and $W_T$ describes a Brownian process.
Let us recall that a Brownian process $W_T$ is a continuous stochastic evolution starting at $W_0=0$ and made of independent gaussian increments. To be specific, let $\mathcal{N}(\mu, \sigma_s)$ be a normal distribution with mean $\mu$ and standard deviation $\sigma_s$. Then, the increment related to two steps of the Brownian processes is  $W_T - W_S \sim \mathcal{N}(0, T - S)$, for $T > S$.

The above differential equation can be solved analytically up to first order using Ito's lemma \cite{ItoLemma-1944}, whereby $W_T$ is treated as an independent variable with the property that $({\rm d}W_T)^2$ is of the order of ${\rm d}T$. Thus, the approximated derivative  ${\rm d}S_T$ can be written as
\begin{equation}
    {\rm d}S_T = \left( \frac{\partial S_T}{\partial T} + \frac{1}{2}\frac{\partial^2 S_T}{\partial W_T^2}\right) {\rm d}T + \frac{\partial S_T}{\partial W_T} {\rm d}W_T.
\end{equation}
By direct comparison to Eq. \eqref{eq:ap_BSM}, it is straightforward to see that
\begin{eqnarray}
    \frac{\partial S_T}{\partial W_T} = S_T\, \sigma, \\ 
    \frac{\partial S_T}{\partial T} + \frac{1}{2}\frac{\partial^2 S_T}{\partial W_T^2} = S_T\, r .
\end{eqnarray}
Using the initial condition $S_0$ at $T=0$, and the Ansatz
\begin{equation}
    S_T = S_0 \exp{(f(T) + g(W_T))},
\end{equation}
the solution for the asset price turns out to be
\begin{equation}
    S_T = S_0 e^{(r - \frac{\sigma^2}{2}) T} e^{\sigma W_T}\;\sim\; S_0 e^{\mathcal{N}\left(\left(r - \frac{\sigma^2}{2}\right) T, \sigma \sqrt{T}\right)}.
\end{equation}
This final result corresponds to a log-normal distribution.

\subsection{European Option}

An option is a contract where in its call/put form, the option holder can buy/sell an asset before a specific date or decline such a right. As a particular case, 
European options can be exercised only on the specified future date, and only depend on the price of the asset at that time. The price that will be paid for the asset is called \emph{exercise} price or \emph{strike}. The day on which the option can be exercised is called \emph{maturity date}. 

A European option payoff is defined as
\begin{equation}\label{eq:Payoff}
    f(S_T, K) = \max(0, S_T - K),
\end{equation}
where $K$ is the strike price and $T$ is the maturity date. An analytical solution exists for the payoff of this kind of options. 

The expected payoff is given by
\begin{equation}
    C(S_T, K) = {\rm average}_{S_T \geq K}\left( S_T - K \right) = \int_{d_1}^{\infty} \left(S_T - K\right) \frac{1}{\sqrt{2\pi}} e^{\frac{-x^2}{2}} dx,
\end{equation}
yielding the analytical solution
\begin{equation}\label{eq:exp_payoff_analytical}
    C(S_T, K) = S_0 {\rm CDF}_{\mathcal N}(d_1) - K e^{-r T}{\rm CDF}_{\mathcal N}(d_2), 
\end{equation}
with 
\begin{center}
\begin{eqnarray}
    d_1 = \frac{1}{\sigma \sqrt{t}}\left( \log \frac{S_0}{K} + \left( r + \frac{\sigma^2}{2}\right) T \right) \\
    d_2 = d_1 - \sigma \sqrt{T} \\
    {\rm CDF}_{\mathcal N}(x) = \frac{1}{\sqrt{2\pi}} \int_{-\infty}^x e^{\frac{-u^2}{2}}du .
\end{eqnarray}
\end{center}

\section{Details for the binary algorithm \label{sec:ap_binary}}

For the sake of completeness, we now present a binary algorithm for option pricing, as introduced in Ref. \cite{qfinance-stamatopoulos2019}. The binary algorithm is also divided in three parts, namely (a) amplitude distribution loading, (b) expected payoff computation and (c) Amplitude Estimation. The main difference is that all computational-basis states are used to codify the discretized probability distribution of an asset price at maturity time. This implies steps (a) and (b) will require completely different quantum circuitry. We now proceed to describe these steps for the binary case.

\subsection{Amplitude distribution loading}

Uploading probability distributions onto quantum states is a very general problem that was considered  in \cite{prob_distributions-grover2002}. In this work, it was claimed that any probability distribution that is efficiently integrable on a classical computer, \textit{e.g.} log-concave distributions, can be loaded efficiently onto a quantum state. However, several authors \cite{finite-montanaro2016, qinspired-garcia2019} have pointed out that the method proposed there requires pre-calculating a number of integrals that grows exponentially with the number of qubits. In the case of option pricing, a reasonable precision requires a moderate number of qubits in the unary representation and many less in the binary representation. But, as a matter of fact, the reduction to a logarithmic number of qubits in the binary representation is, at least partially, compensated by the effort needed to prepare the probability distribution.
That is, in practise, both unary and binary representations require similar effort to pre-process the probability distribution to later encode it in the quantum register. 

An alternative method to encode a probability distribution in a quantum state is the use of so-called quantum Generative Adversarial Networks (qGANs) \cite{qGAN-lloyd2018, qGAN-dallaire2018, qGAN-zoufal2019}. In this scheme, two agents, a generator and a discriminator compete against each other. The generator learns to produce data that mimics the underlying probability distribution, trying to deceive the discriminator into believing that the new data  is faithful. On the other hand, the discriminator has to learn how to tell apart the real data from the  data produced by the generator. This quantum adversarial game has a unique endpoint: Nash equilibrium is reached when the generator learns to produce states that deliver probability outcomes that are indistinguishable from the desired probability distribution, and the discriminator cannot tell them apart. In order to upload probability distributions onto quantum states using qGANs, a parametrized quantum circuit may play the role of a generator, whereas the discriminator may be a classical neural network.

At present, there is still a lack  of precise understanding on how to efficiently upload probability distributions on a quantum computer in binary representation, which makes rigorous complexity analysis in terms of the number of gates difficult.

\subsection{Payoff computation}\label{subsec:ap_payoff_binary}

\begin{figure}[ht]
    \centering
    \includegraphics[width=1.2\linewidth]{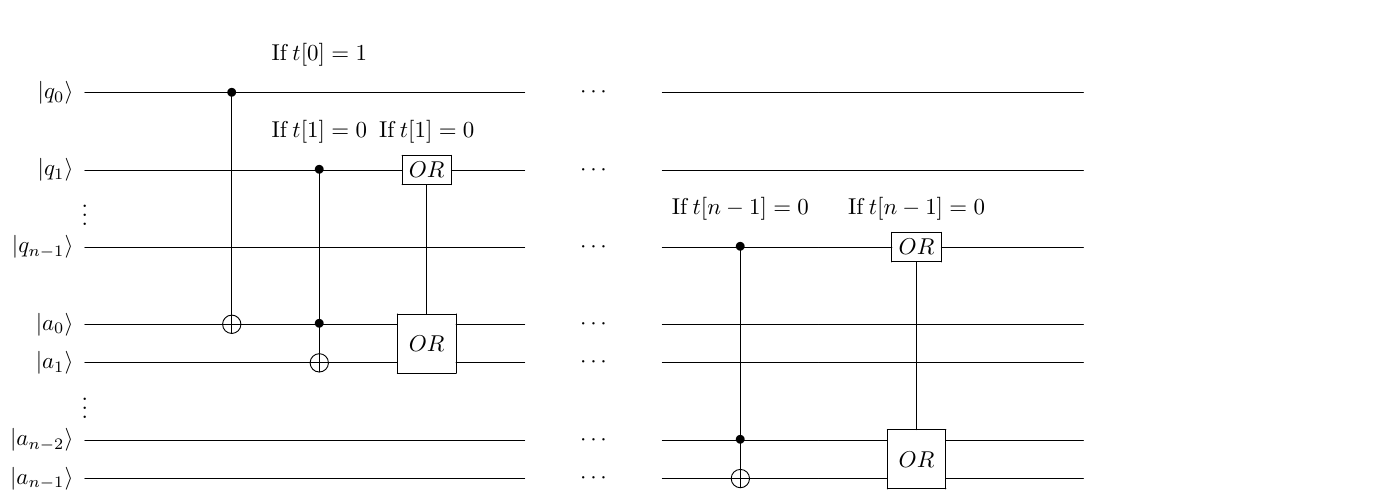}
    \caption{Quantum comparator, $\C$. The OR gates appearing are three-qubit gates acting on non-adjacent qubits.}
    \label{fig:comparator}
\end{figure}

A useful feature of the unary algorithm is that, given a strike $K$, one can directly know which qubits will not contribute to the expected return of the option, and therefore adjust the quantum circuit. This is only possible since the unary representation maps directly to the asset price. In a binary encoded setting one needs to compute explicitly which basis elements will make a non-zero contribution to the expected payoff. Hence the need of a quantum comparator, $\C$, that singles out the values of $S_T$ that are smaller than the strike price $K$. This comparator requires the use of $n+1$ ancillary qubits, one of which is retained after the computation.  Its action is given by
\begin{equation}
    \label{eq:ap_C} |\psi\ra|0\ra\xrightarrow{\quad \C \quad} \sum_{S_i<K} \sqrt{p_i}\,|e_i\ra|0\ra + \sum_{S_i\geq K} \sqrt{p_i}\,|e_i\ra|1\ra,
\end{equation}
where  $\{|e_i\ra\}$ is the computational basis and $\{S_i\}$ are the asset values at maturity associated to computational-basis vectors. The quantum circuit implementing Eq. \eqref{eq:ap_C} can be constructed using cNOTs, Toffoli gates and  OR gates, see Fig. \ref{fig:comparator}. 
In order to understand the way  this circuit works, let us consider the case where the discretization of the interval $[S_{max}-S_{min}]$ is uniform. In this case, the relation between $\{S_i\}$ and $\{|e_i\ra\}$ is
\beq S_i = S_{min}+ \frac{e_i\,(S_{max}-S_{min})}{2^n} .\eeq 
This implies that
\beq \label{eq:k'} S_i > K \quad \Leftrightarrow \quad e_i >  \frac{2^n\,(K-S_{min})}{S_{max}-S_{min}} \equiv K'.\eeq
The idea goes as follows. First, we classically compute the two's complement of $K'$, \ie $2^n-K'$, and store it in binary format in a classical array of $n$ bits, $t[j]$ with $j\in[0,1,\dots,n-1]$. Then, using $n$ ancillas, $|a_0\cdots a_{n-1}\ra$, initialized to $|0\cdots0\ra$, we compute the carry bits of the bitwise addition between $t$ and $\{e_i\}$, and store them in superposition into $|a_0\cdots a_{n-1}\ra$. If $e_i>K'$, then necessarily $a_{n-1}=1$.

\begin{figure}[ht]
    \centering
    \includegraphics{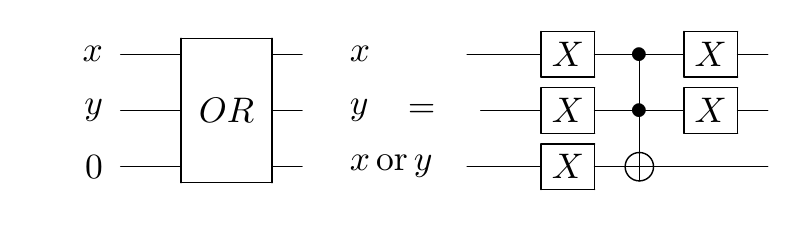}
    \caption{Decomposition of the OR gate in terms of single-qubit and Toffoli gates.}
    \label{fig:qOR}
\end{figure}

The exact circuit needed for a given strike will depend upon the values of the bits in $t$. If $t[j]=0$, then there will be a carry bit at position $j$ if and only if there is a carry bit at position $j-1$ {\sl and} the $j$-th bit of $e_i$ is 1. This is computed with a Toffoli gate. On the other hand, if $t[j]=1$, there will be a carry bit at position $j$ if and only if there is a carry bit at position $j-1$ {\sl or} the $j$-th bit of $e_i$ is 1. This is computed with an OR gate, shown in Fig.\ref{fig:qOR}. Finally, there will be a carry bit at $a_0$ if and only if $t[0]=1$ and the first bit of $e_i$ is 1. This is achieved with a simple CNOT gate. As explained above, if $e_i\geq K'$, then $a_{n-1}$ must be equal to 1. Hence, applying a CNOT gate controlled by the qubit $\ket{a_{n-1}}$ and targeted at the ancilla, the desired state in Eq. \eqref{eq:ap_C} is obtained.

Once $\C$ has been applied, the next step is to encode the expected payoff of the option into the amplitudes of a new ancilla. The final state to be created should be
\beq \label{AE} \sum_{S_i<K} \sqrt{p_i}\,|e_i\ra|0\ra\left[\cos(g_0)|0\ra+\sin(g_0)|1\ra\right] + \sum_{S_i\geq K} \sqrt{p_i}\,|e_i\ra|1\ra \left[\cos(g_0+g(i))|0\ra+\sin(g_0+g(i))|1\ra\right],\eeq
where
\beq g_0=\frac{\pi}{4}-c \qquad,\qquad g(i)=\frac{2c\,(e_i-K')}{e_{max}-K'}\,, \eeq
with $c$ a constant such that $c\in [0,1]$. Thus, the probability of measuring the second ancilla in the $|1\ra$ state in \eqref{AE} is
\beq {\rm{Prob}}(1)=\sum_{S_i<K} p_i \sin^2(g_0)+\sum_{S_i\geq K} p_i \sin^2(g_0+g(i)) \,.\eeq
Using the approximation
\beq \label{approx} \sin^2\left(cf(i)+\frac{\pi}{4}\right)=\frac{1}{2}+cf(i)+O(c^3f^3(i))\,,\eeq
to first order, which follows from Taylor-expanding $\sin^2(f(x)+\frac{\pi}{4})$ around $f(x)=0$, the probability becomes
\beq \label{P1} {\rm{Prob}}(1) \simeq \sum_{S_i<K} p_i \left(\frac{1}{2}-c\right)+\sum_{S_i\geq K} p_i \left(\frac{1}{2} +c\left[\frac{2\,(e_i-K')}{e_{max}-K'} -1\right]\right) = \frac{1}{2}-c+\frac{2c}{e_{max}-K'} \sum_{S_i\geq K}\, p_i \,(e_i-K') \,.\eeq 
It is important to note that the approximation made in Eq. \eqref{P1} is valid since $cf(i)= c\left[\frac{2\,(e_i-K')}{e_{max}-K'} -1\right] \in[-c,c]$.
Reversing the change from Eq. \eqref{eq:k'}, namely
\begin{equation}
    \sum_{S_i\geq K} p_i \,(S_i-K) = \frac{S_{max}-S_{min}}{2^n}\sum_{S_i\geq K}\, p_i \,(e_i-K'),
\end{equation}
the expected payoff function, \ie $\sum_{S_i\geq K} p_i \,(S_i-K)$, can be recovered from the probability of measuring 1 in the ancilla, Eq. \eqref{P1}, since $c,K,S_{max}$ are all known. 
\begin{figure}[h!]
    \centering
\includegraphics{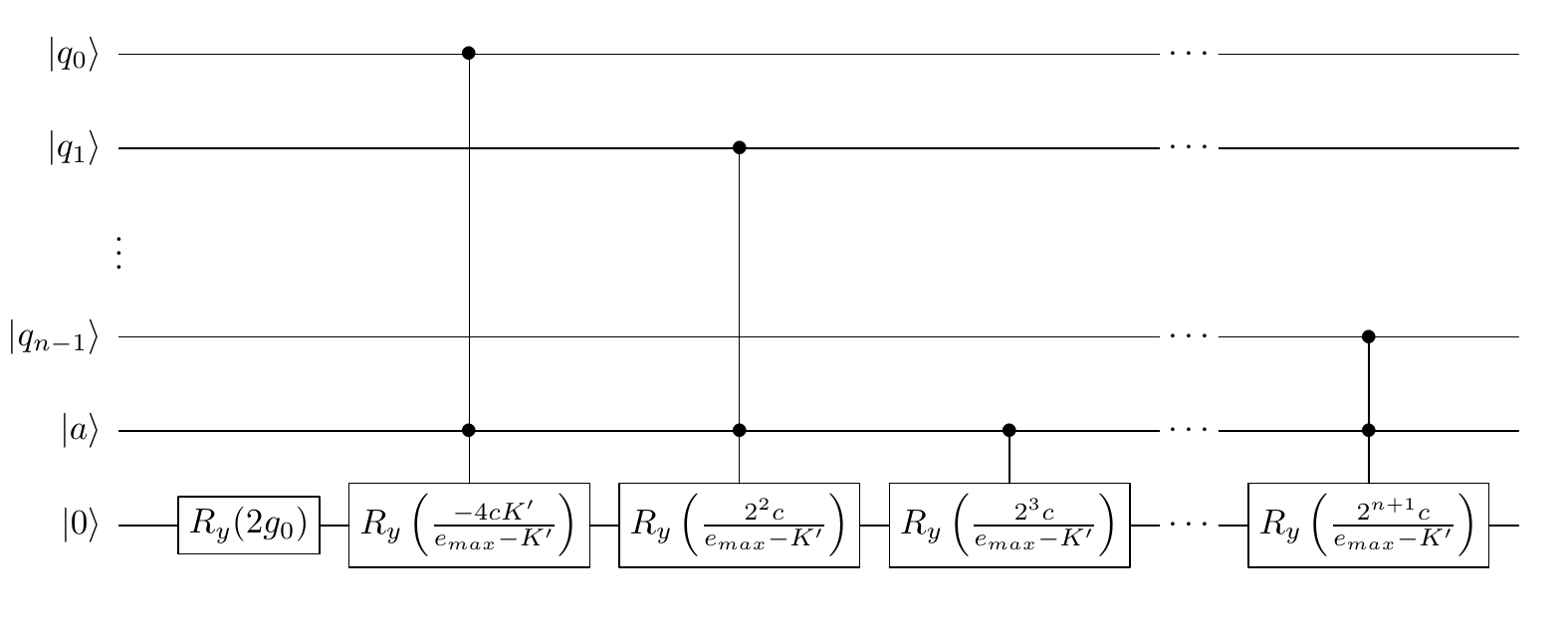}
    \caption{Encoding of the expected return of a European option into the amplitudes of an ancilla qubit in binary representation.}
    \label{fig:binary_encoding}
\end{figure}
\begin{figure}[h!]
	\centering
	\includegraphics{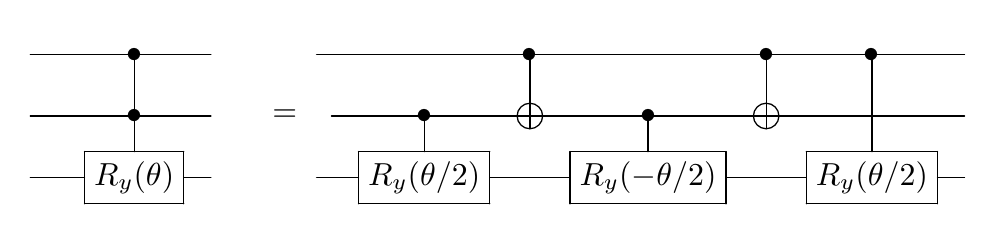}
	\caption{\small{Decomposition of a cc$R_y$ gate in terms of cNOTs and c$R_y$ gates.}}
	\label{fig:ccRy}
\end{figure}

The quantum circuit that produces the final state \eqref{AE} from Eq. \eqref{eq:ap_C} is composed of $n$ cc$R_y$ gates, plus one $R_y$ and one c$R_y$ gate, shown in Fig. \ref{fig:binary_encoding}. The decomposition of a cc$R_y$ gate in terms of CNOTs and c$Ry$ gates is shown in Fig. \ref{fig:ccRy}.

\subsection{Amplitude Estimation}

The oracle operator $\S_{\psi_0}$ acts on the binary algorithm in the same manner as its unary analogue, as defined in Eq. \eqref{eq:oracle}. 

The case of the operator $\S_0$ is slightly different. The target state to flip is that with only $\ket 0$. Thus, a multi-controlled Toffoli gate is required. This multi-controlled gate constitutes the computationally most-costly piece of the circuit. This gate can be decomposed in simpler gates \cite{gates-barenco1995}.

In order to reduce the complexity of the circuit as much as possible, it is necessary to choose the optimal representation of this gate. The most efficient decomposition consists in a chain of standard Toffoli gates that store their outcomes in ancilla qubits. The number of ancillas required is $c - 2$, where $c$ is the number of control qubits. Depending on whether these ancillas are initialized in $0$ or not, the number of Toffoli gates is different. In the case of the binary algorithm, a sufficient number of ancillas is available from the comparator. In addition, the short version of the Toffoli gate can be used. The reason is as follows. The operator $\S_0$ is applied only after the sequence $\A^\dagger \S_{\psi_0} \A$, that leaves the ancillas unchanged as the oracle includes only a phase and does not add any entanglement, making all operations involving the ancillas classical. In other words, there is no need to add any circuit piece that undoes the auxiliary computations stored in the ancillary qubits because the structure of the circuit itself accomplishes this goal. 

\begin{figure*}
    \centering
    \includegraphics{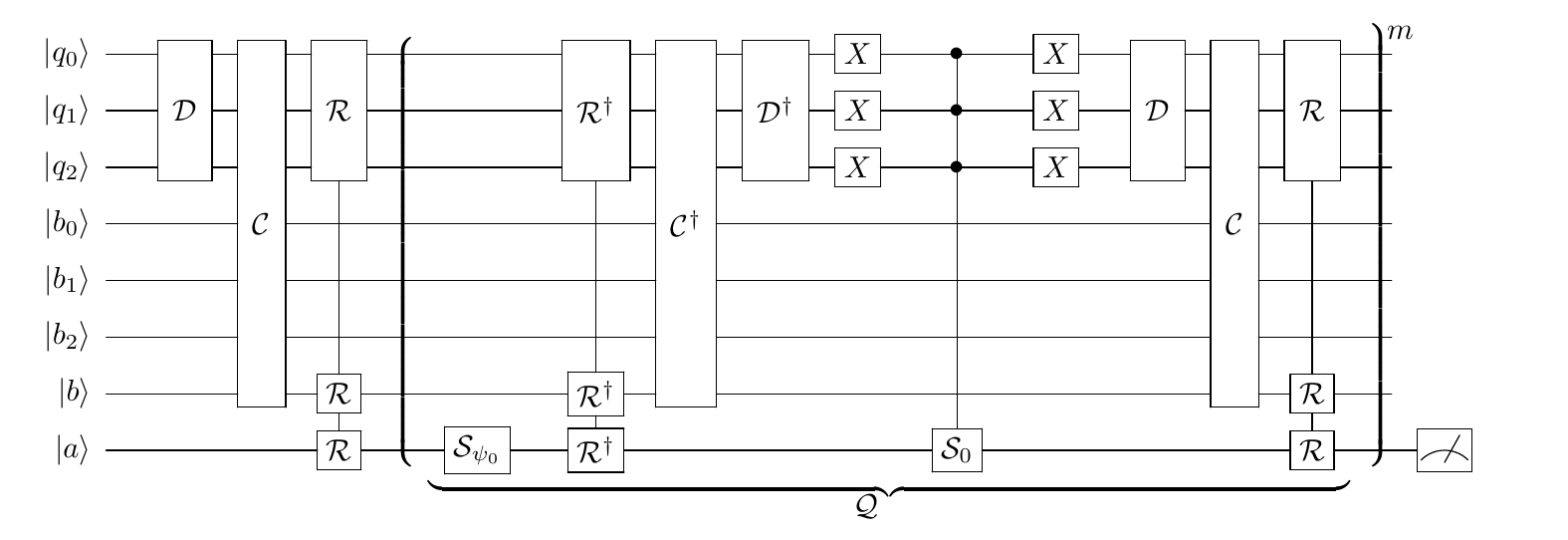}
    \caption{Full circuit for the option pricing algorithm in the binary representation. The gate $\D$ is the probability distributor, $\C$ is the comparator, and $\R$ the rotation step. $\C$ and $\R$ together represent the computation of the payoff. After applying the algorithm, the oracle $\S_{\psi_0}$, the reverse algorithm and $\S_0$ follow. The last step is applying the algorithm again. This block $\Q$ is to be repeated for applying Amplitude Estimation. }
    \label{fig:full_binary}
\end{figure*}

The full circuit for the binary algorithm, including Amplitude Estimation is depicted in Fig. \ref{fig:full_binary}

\section{Details for the Amplitude Distributor in the unary basis \label{sec:ap_amplitude_distribution}}

Let us consider Fig. \ref{fig:ProbUploading}. In the unary basis, every qubit represents the basis element in which the qubit is $\ket{1}$. Thus, the coefficient of every element depends on as many angles as partial-SWAP gates are needed to reaching its corresponding qubit. Thus, the central qubits of the circuit will depend only on 2 angles, and the number of dependencies increases one by one as we move to the outer part of the circuit. The very last two qubits depend on the same angles. As we move away from the middle, each qubit inherits the same angle dependency than the previous ones plus an additional rotation.
Starting from the two edges, their coefficients verify the following ratios
\begin{eqnarray}
    \left\vert\frac{\psi_{0}}{\psi_{1}}\right\vert ^2 & = & \tan^2(\theta_{1} / 2) \\
    \left\vert\frac{\psi_{n-1}}{\psi_{n-2}}\right\vert ^2 & = & \tan^2(\theta_{n - 1} / 2).
\end{eqnarray}
Then we identify $|\psi_i|^2 = p_i$, where $\lbrace p_i\rbrace$ is the target probability distribution of the asset prices at maturity. The next step corresponds to considering the qubits $1$ and $2$, as well as $n-3$, $n-2$. The relations for their coefficients must obey
\begin{eqnarray}
    \left\vert\frac{\psi_{i}}{\psi_{i+1}}\right\vert ^2 & = & \cos^2(\theta_i/2)\tan^2(\theta_{i+1} / 2) \\
    \left\vert\frac{\psi_{n-1-i}}{\psi_{n-2-i}}\right\vert ^2 & = & \cos^2(\theta_{n-i}/2)\tan^2(\theta_{n - 1 - i} / 2).
\end{eqnarray}
Then, it is straightforward to back-substitute parameters step by step until we arrive to the central qubits.
This procedure fixes all the angles for the partial-SWAP gates used in the amplitude distributor.

The exact algorithm to be followed can be also found in the provided code \cite{github}.

Once the exact solution for the angles is inserted into the circuit depicted in Fig. \ref{fig:ProbUploading}, the amplitude distributor algorithm is completed. The quantum register then reads
\begin{equation}
    \ket{\Psi}=\sum^{n-1}_{i=0}\sqrt{p_i}\ket{i}.
\end{equation}
Note that describing a probability distribution with squared amplitudes of a quantum state allows for a free phase in every coefficient of the quantum circuit. For simplicity, we will set to zero all these relative phases by only operating with real valued partial-SWAP gates.

Let us turn our attention to the gates which are needed in the above circuit.
Sharing probability between neighbor qubits can be achieved by introducing a two-qubit gate based on the SWAP and $R_y$ operations. This variant on the SWAP gate performs a partial SWAP operation, where only a piece of the amplitude is transferred from one qubit to another. This operation preserves unarity, that is the state remains as a superposition of elements of the unary basis. This partial-SWAP, can be decomposed using CNOT as the basic entangling gate as 
\begin{equation}\label{eq:ap_SWAPRy}
      \includegraphics[width=0.175\linewidth, valign=c]{SWAPRy2.png} \quad=
    \left(\begin{array}{cccc}
    1&0&0&0 \\ 0&\cos{\theta/2}&\sin{\theta/2}&0 \\0&-\sin{\theta/2}&\cos{\theta/2}&0 \\0&0&0&1
    \end{array}\right)=\quad
    \quad\includegraphics[width=0.2\linewidth, valign=c]{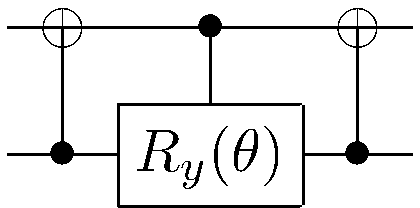},
\end{equation}
where the usual CNOT gate in the center of the conventional SWAP gate has been substituted by a controlled $y$-rotation, henceforth referred to as c$R_y$ gate. In turn, the c$R_y$ operation can be reworked as a combination of single-qubit gates and CNOT gates \cite{gates-barenco1995}:
\begin{equation}\label{CRy}
    \includegraphics[width=0.15\linewidth, valign=c]{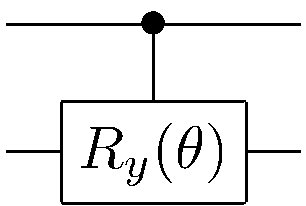} \quad=
    \quad\includegraphics[width=0.325\linewidth, valign=c]{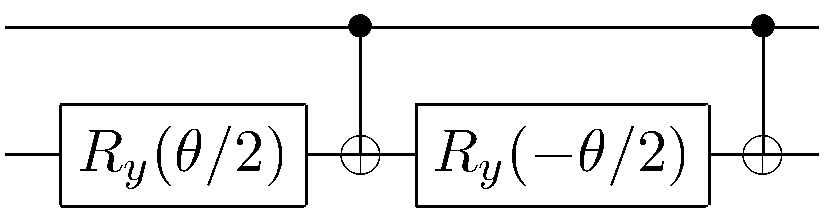}.
\end{equation}
This decomposition will come into play for the expected payoff calculation algorithm as well, albeit with angle $\phi$ in the payoff circuit.

For the purposes of this algorithm, both the CNOT and partial-iSWAP basis gates are analogous, but the direct modeling to partial-iSWAPs can economize the total number of required gates for the amplitude distributor. Partial-iSWAP gates can be used to decompose CNOT gates. More explicitly, a CNOT gate an be reproduced with two iSWAP gates, and 5 single qubit gates.

\begin{algorithm}[t!]
\caption{\label{alg:gaussian} Algorithm for Amplitude Estimation based on gaussian distribution of the measurements. }
\DontPrintSemicolon
\SetKwFunction{FMain}{GaussianAmplitudeEstimation}
\SetKwFunction{FArcSin}{MultipleValuesArcsin}
\SetKwProg{Fn}{Function}{:}
\Fn{\FMain{$N_{\rm shots}$, $J$, $m_j$, $\alpha$}}
{\;
    $z \leftarrow {\rm CDF}_\mathcal{N}^{-1}(1 - \alpha / 2)$ \;
    Ensure $m_0 = 0 $ \;
    $a \leftarrow |\bra{1}\A\ket 0 |^2$ with $N_{\rm shots}$ samples\;
    $\theta_a^{(0)} \leftarrow \arcsin{\sqrt{a}}$
    $\Delta \theta_a^{(0)} = \frac{z}{2 \sqrt{N_{\rm shots}}}$\;
    \For{$j \leftarrow 1 \KwTo \; J$}
    {
        $a \leftarrow |\bra{1}Q^{m_j}A\ket 0 |^2$ with $N_{\rm shots}$ samples\;
        $\theta_{\rm array} \leftarrow $ \FArcSin{$a, m_{j-1}$} \;
        $\theta_a \leftarrow \min\left(|\theta_{\rm array} - \theta_a^{(j - 1)}|\right)$\;
        $\Delta \theta_a \leftarrow \frac{z}{2 (2 m_j + 1)\sqrt{N_{\rm shots}}}$\;
        $\theta_a^{(m_j)} \leftarrow \frac{\frac{\theta_a }{\Delta \theta_a^2} + \frac{\theta_a^{(j-1)} }{(\Delta \theta_a^{(j-1)})^2}}{\frac{1}{\Delta \theta_a^2} + \frac{1}{(\Delta \theta_a^{(j-1)})^2}}$\;
        $\Delta \theta_a^{(m_j)} \leftarrow \left(\frac{1}{\Delta \theta_a^2} + \frac{1}{(\Delta \theta_a^{(j-1)})^2}\right)^{-1/2}$\;
        $[a_j, \Delta a_j] \leftarrow [\sin^2\theta_a^{j}, \sin(2 \theta_a^{j}) \Delta \theta_a^{(j)}]$
    }
    \KwRet{$[a_j, \Delta a_j]$} }
\end{algorithm}

\begin{algorithm}[t!]
\caption{\label{alg:gaussian2} Extracting multiple values for the $\arcsin$, auxiliary function needed in Alg. \ref{alg:gaussian}.}
\DontPrintSemicolon
\SetKwFunction{FArcSin}{MultipleValuesArcsin}
\SetKwProg{Fn}{Function}{:}
\Fn{\FArcSin{$a, m$}}
{\;
    $\theta_0 \leftarrow \arcsin{\sqrt{a}}$ \tcp{The value of $\theta_0$ is bounded between $0$ and $\pi / 2$}
    The $\arcsin$ function has several solutions
    $\theta_{\rm array} \leftarrow [0] * (2 m + 1)$
    $\theta_{\rm array}[0] \leftarrow \theta_0$\;
    \For{$k \leftarrow 1 \KwTo\, m$}
    {
        $\theta_{\rm array}[2k - 1]\leftarrow k \pi - \theta_0$\;
        $\theta_{\rm array}[2k]\leftarrow k \pi + \theta_0$
    }
    $\theta_{\rm array} \leftarrow \theta_{\rm array} / (2 m + 1)$\;
    \KwRet $\theta_{\rm array}$}
\end{algorithm}

\section{Selection of results for the Iterative Amplitude Estimation\label{sec:ap_iae}}

We present here a method for obtaining the most probable value of $a$ in an iterative fashion following similar methods as other Amplitude Estimation without QPE algorithms. We base this procedure in the theory of confidence intervals for a binomial distribution assuming normal distributions \cite{binomial-wallis2013}.

Let us consider a binomial distribution with probability $a$, \ie for every sample the chance of obtaining $1$ is $a$, while the chance of obtaining $0$ is $1 - a$. Then, if an estimate $\hat a$ of $a$ was obtained using $N$ samples, the true value of $a$ lies in the interval
\begin{equation}
    a = \hat a \pm \frac{{\rm CDF}_\mathcal{N}^{-1}(1 - \alpha/2) \sqrt{\hat a(1-\hat a)}}{2 \sqrt{N}}, 
\end{equation}
with confidence $(1 - \alpha)$.

From this result we can construct an iterative algorithm returning the optimal value of $a$ using quantum Amplitude Estimation. Let us take a set of $m_j$ for $j={0,1,2,3,\ldots}$. For every $m_j$ the probability of obtaining $\ket 1$ is $\sin^2((2 m_j + 1) \theta_a)$, where $a = \sin^2(\theta_a)$. Let us move to the $\theta$ space. For a given $m$ the values and error of $\theta$ obtained are
\begin{equation}\label{eq:weighted}
    \theta_a = \frac{1}{2m+1}\arcsin(\sqrt{a}) \qquad \Delta \theta_a = \frac{1}{2m + 1}\frac{{\rm CDF}_\mathcal{N}^{-1}(1 - \alpha/2)}{2 \sqrt{N}}.
\end{equation}
It is important to understand two main properties of Eq. \eqref{eq:weighted}. First, there are $2m + 1$ possible values for $\theta_a$ within the interval $\theta_a \in [0, \pi / 2]$ as the $\sin^2(\cdot)$ function is $\pi$-periodical. For every new iteration it will be necessary to choose one of them.  It is very important to set $m_j = 0$ at first because this case is the only one for which $\theta_a$ corresponds to the expected value for $a$. Otherwise, several possible values of $a$ arise and it is not possible to tell which one is correct. Combining results for several values of $m_j$, it is possible to bound the uncertainty to be as small as desired.

The algorithm is based on the following statements. 
For a given collection of measurements and uncertainties $\{\theta_i, \Delta \theta_i\}$, the weighted average and uncertainty from the first $j$ terms is
\begin{equation}\label{eq:ap_exp_errors}
    \Tilde{\theta}_j = \frac{\sum_{i=0}^j \theta_i / \Delta \theta_i^2}{\sum_{i=0}^j 1 / \Delta \theta_i^2}  \qquad \Delta \Tilde{\theta}_j = \left(\sum_{i=0}^j 1 / \Delta \theta_i^2\right)^{-1/2}.
\end{equation}
Notice also that this relation is recursive, as $\Tilde{\theta}_{j+1}$ can be obtained by combining $\Tilde{\theta}_{j}$ and $\theta_{j+1}$. The same holds for uncertainties. Thus, the interpretation of this algorithm is that for every new step $j$ a new term is added to the series $\{\theta, \Delta \theta\}$. The individual uncertainties decrease as $\sim ((2 m + 1)^{-1})$, and the final global uncertainty is obtained as
\begin{equation}\label{eq:uncertainty}
    \Delta \theta = \frac{{\rm CDF}_\mathcal{N}^{-1}(1 - \alpha / 2)}{\sqrt{N}}\left( \sum_{j=0}^J (2m_j + 1)^2 \right)^{-1/2},
\end{equation}
where $J$ denotes the last iteration performed. The full recipe for the algorithm is described in Algs. \ref{alg:gaussian} and \ref{alg:gaussian2}.

In the case of a linear selection of $m_j$, \ie $m_j = j; j=(0, 1, 2, ..., J)$, the asymptotic behavior of this uncertainty is $\Delta \theta = \mathcal{O}(N^{-1/2} M^{-3/4})$, with $M$ the sum of all $m$. For discovering it we just have to compute

\begin{equation}
    \sum_{j=0}^J (j + 1)^2 = 4 \sum_{j=0}^J j^2 + 4 \sum_{j=0}^J j + \sum_{j=0}^J 1.
\end{equation}
We now take the identities $\sum_{j=0}^J j = J(J+1)/2=M$ and $\sum_{j=0}^J j^2 = J(2J+1)(J+2)/6$. Then, it is direct to check that
\begin{equation}\label{eq:prec_lineal}
    \Delta \theta = \mathcal{O}(N^{-1/2} J^{-3/2}) = \mathcal{O}(N^{-1/2} M^{-3/4}).
\end{equation}

This behavior already surpasses the tendency of the classical sampling, but does not reach the optimal Amplitude Estimation.

In the case of an exponential selection of $m_j$, \ie $m_j = \{0\} \cup \{2^j\}; j=(0, 1, 2, ..., J)$ we can take the identities $\sum_{j=0}^J 2^j = 2^J - 1 =M$ and $\sum_{j=0}^J 2^{2j} = (2^{2J} - 1) / 3$. Then it is direct to check that
\begin{equation}\label{eq:prec_exp}
    \Delta \theta = \mathcal{O}(N^{-1/2} 2^{-J}) = \mathcal{O}(N^{-1/2} M^{-1}).
\end{equation}

\subsection{Extension to error-mitigation techniques}

The error-mitigation procedure proposed for the unary algorithm discards some of the algorithm instances to retain outcomes within the unary basis. This reduces the precision achieved in the algorithm with respect to the ones predicted in Eqs. \eqref{eq:prec_lineal} and \eqref{eq:prec_exp} in order to maintain accuracy. This section provides some lower bounds on how many AE iterations can be done while still reaching quantum advantage. 

We will work now in the scheme where $m_j = j$. Let us assume that, in every iteration of Amplitude Estimation, only a fraction $\tilde p_j$ of the shots are retained. The equivalent version of Eq. \eqref{eq:uncertainty} is now
\begin{equation}
    \Delta \theta = {\rm CDF}_\mathcal{N}^{-1}(1 - \alpha / 2)\left( \sum_{j=0}^J (2m_j + 1)^2 N\tilde p_j \right)^{-1/2}.
\end{equation}

As more errors are bound to occur, $\tilde p_j$ decreases as $m_j$ increases, we can state a bound for the accuracy as
\begin{equation}
    \Delta \theta \leq \frac{{\rm CDF}_\mathcal{N}^{-1}(1 - \alpha / 2)}{\sqrt{N \tilde p_J}}\left( \sum_{j=0}^J (2m_j + 1)^2 \right)^{-1/2}, 
\end{equation}
since the precision is at least as good as the one obtained for the worst-case scenario. Comparing the trends, both in the linear and the exponential case, with the classical scaling, it is possible to see that quantum advantage is still achieved provided
\begin{equation}\label{eq:p_J}
\tilde p_J \geq M^{1 - 2\alpha},    
\end{equation}
with $\alpha = 3/4$ in the linear case and $\alpha = 1$ in the exponential case.

The probability of retaining a shot is at least the probability of having no errors in the circuit, considering that some double errors may lead to erroneous instances but in the unary basis. This zero-error probability in the worst case scenario, that is, at the last iteration of AE, is written as
\begin{equation}
    p_0 = \left((1-p_e)^{a n + b}\right)^{m_J},
\end{equation}
where $p_e$ is the error of an individual gate, and $a$ and $b$ are related to the gate scaling, see Tab. \ref{tab:gates} for the details. In principle, one can expand the calculation of $p_0$ by considering different kinds of errors for different gates, but for the sake of simplicity we will focus on this analysis. Rearranging together the results for Eqs. \eqref{eq:prec_lineal}, \eqref{eq:prec_exp} and \eqref{eq:p_J} it is possible to see that quantum advantage is obtained if the individual gate errors is bounded by
\begin{equation}
    p_e < 1 - m_J^{\frac{2 - 4\alpha}{(a n + b)m_J}}.
\end{equation}

\section{Independent analysis of errors for Amplitude Estimation \label{sec:ap_more_results}}

In this section we present results that extend the ones depicted in Sec. \ref{sec:simulations}.

\begin{figure}[t!]
\centering
    {\includegraphics[width=.45\textwidth]{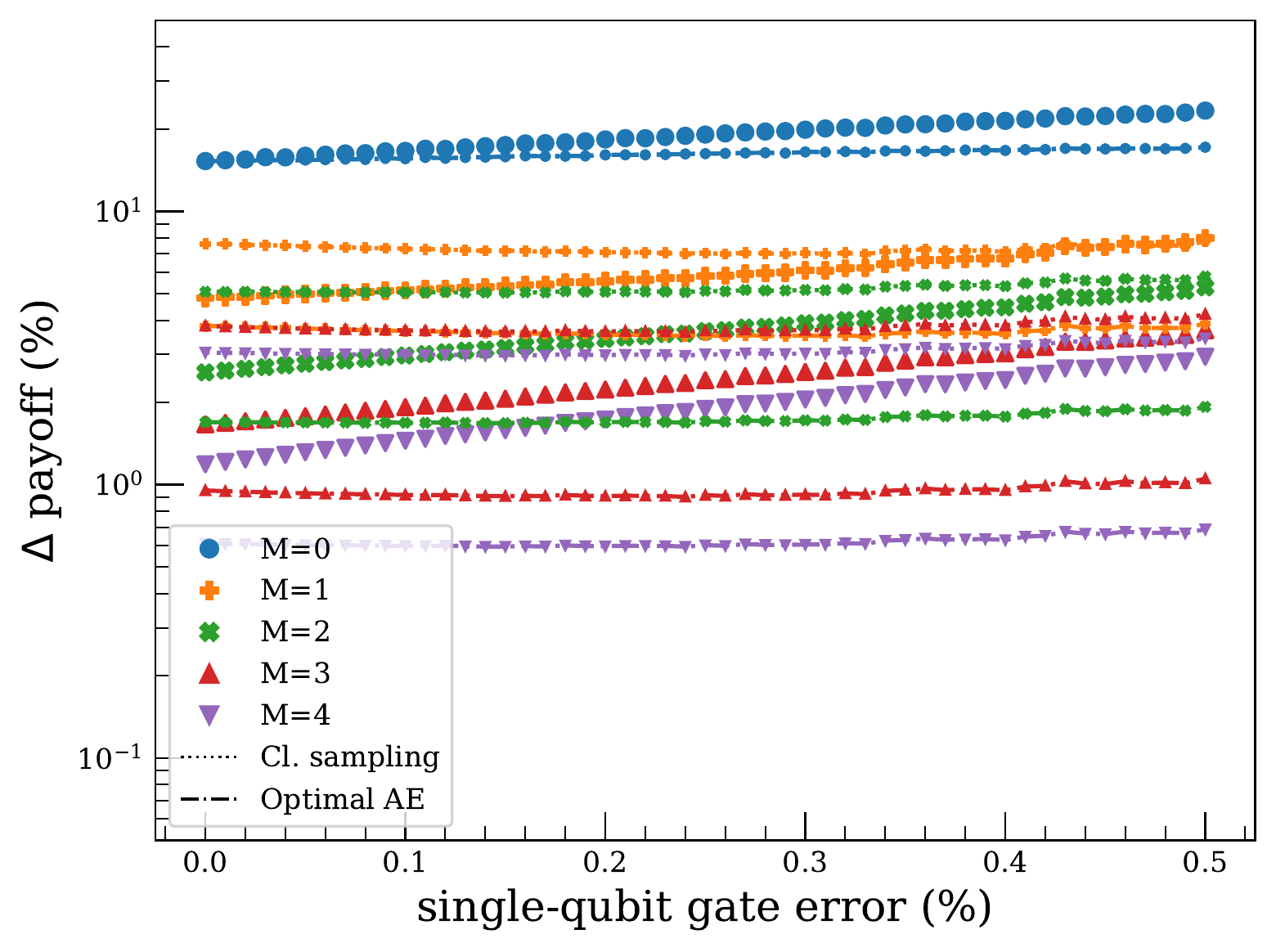}}
    \hfill {\includegraphics[width=.45\textwidth]{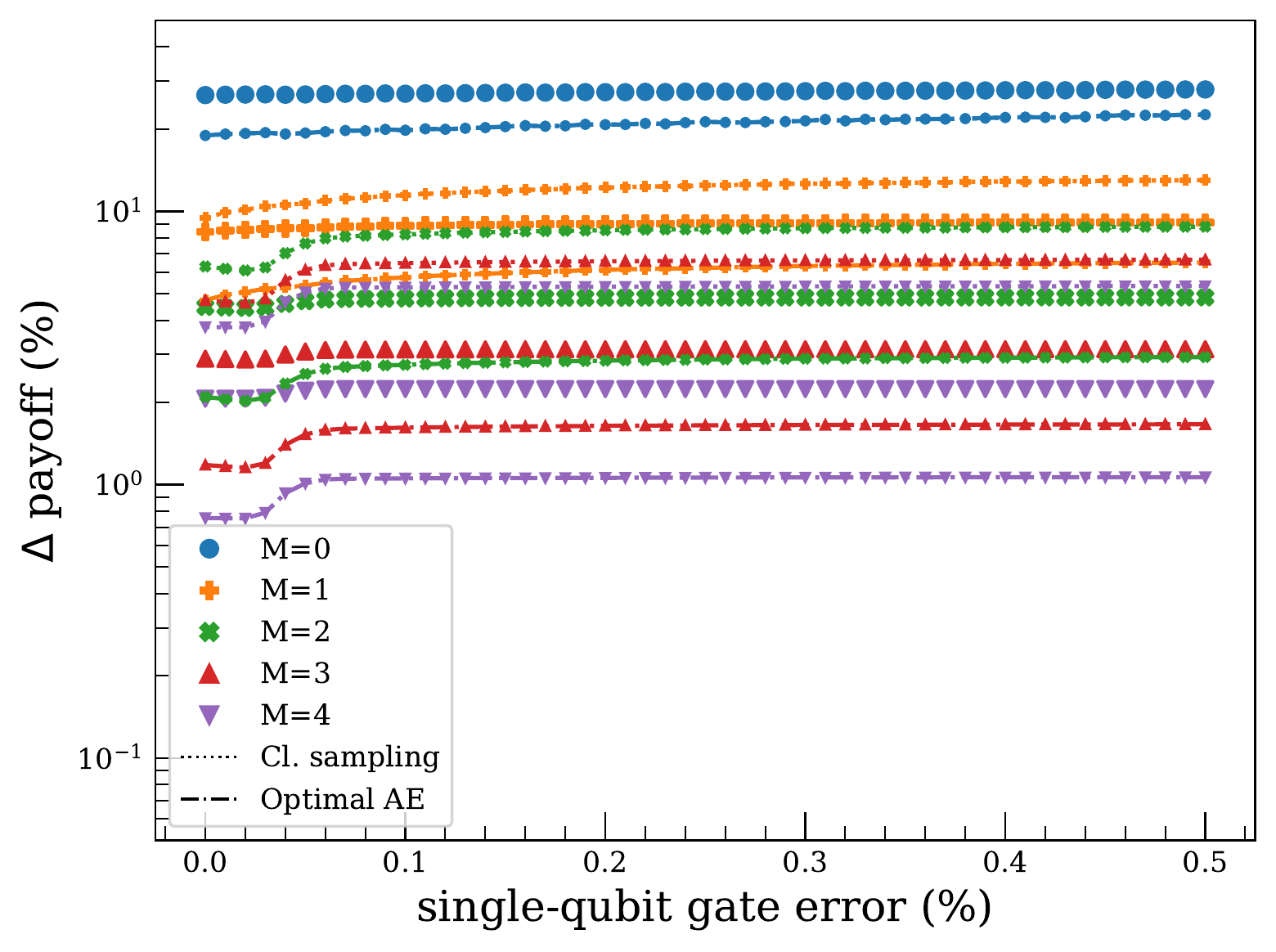}}
    \caption{Results of the sampling uncertainties of the expected payoff, for the unary (Left) and binary (Right) representation, with $M$ iterations of Amplitude Estimation for both the unary and binary approaches, considering depolarizing and read-out errors together. Scattering points represent  the uncertainties obtained for the experiment while dash-point lines represent theoretical bounds, where each line is accompanied with the corresponding marker. For every color and symbol, the lower bound is for optimal quantum advantage, and the upper bound is for sampling. For $M=0$, both bounds are equivalent. In every case, a new iteration of the Amplitude Estimation reduces the uncertainty. For the unary case, the scattering points present a tendency to return larger uncertainties as the errors increase, while for the binary case the uncertainties do not explicitly depend on the single-qubit gate error. This difference is a direct consequence of post-selection.}
    \label{fig:errors}
\end{figure}

It is also interesting to see how the uncertainties of the measurements decrease as more iterations of Amplitude Estimations are allowed. We must remark that those errors are exclusively due to the uncertainty in the sampling. This can be observed in Fig. \ref{fig:errors}, where the obtained uncertainties are bounded between the classical sampling and the optimal Amplitude Estimation.

Furthermore, a very remarkable behavior of the uncertainties in the unary approach must be noticed. The obtained uncertainties present a tendency to increase as errors get larger. In contradistinction, the binary algorithm does not present this feature. The reason lies in the post-selection that is only applicable in the unary representation. As errors become more likely to happen, the post-selection filter rejects more instances. The direct consequence is that the number of accepted shots drops for large errors, causing less certain outcomes. The joint action of this processes is that the uncertainty decreases more slowly for the unary algorithm than for the binary one. This behaviour contrasts with the error obtained in Fig. \ref{fig:data}, where the binary results reflect a very poor performance.

\begin{figure}
    \centering
    \includegraphics[width=.45\textwidth]{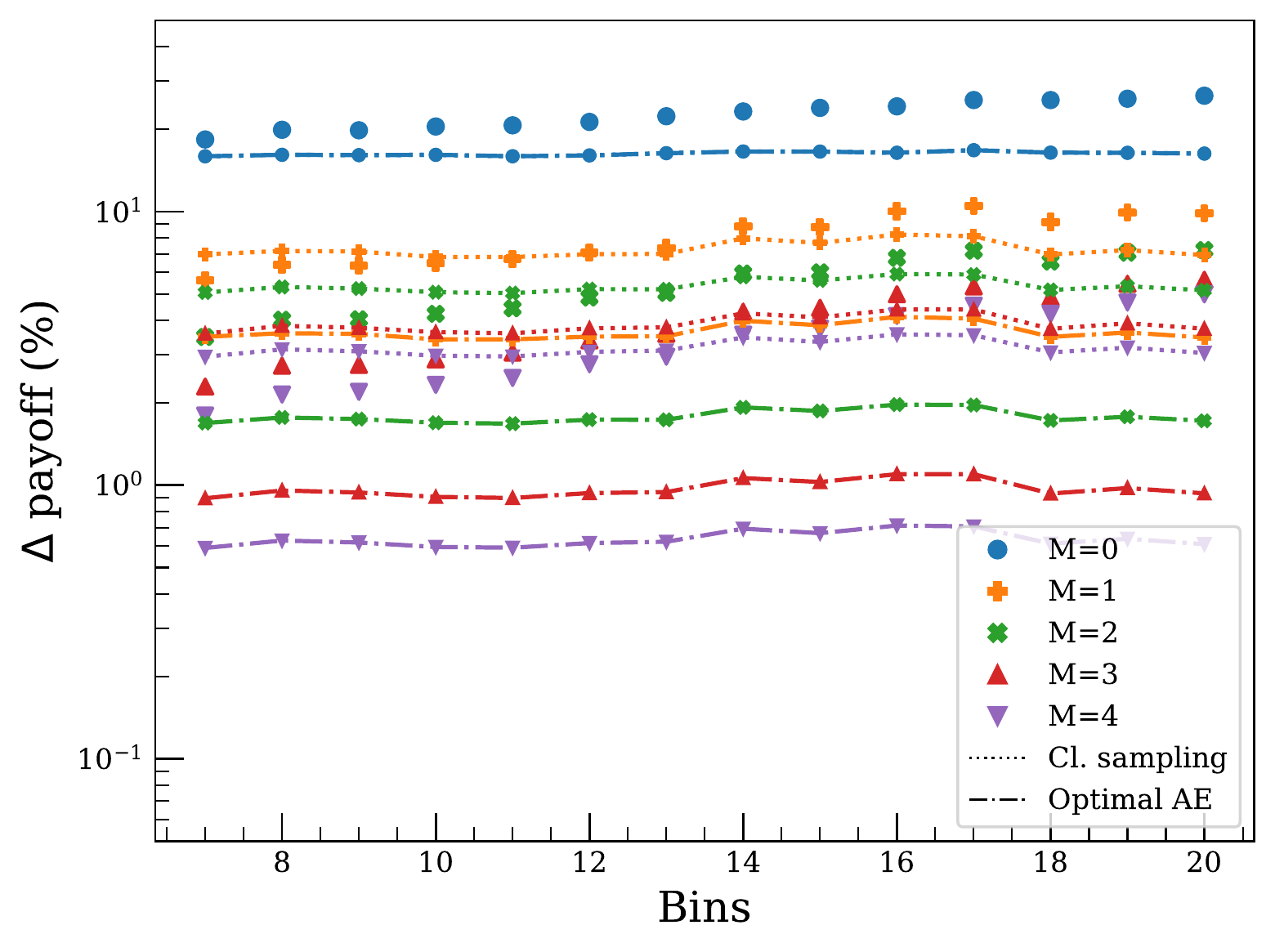}
    \caption{Results for the sampling uncertainties of the expected payoff for increasing number of bins for up to $M$ iterations of Amplitude Estimation for the unary approach, considering depolarizing and read-out errors together. Scattering points represent uncertainties obtained and dash-point lines represent theoretical bounds, where each line is accompanied with the corresponding marker. For every color and symbol, the lower bound is for optimal quantum advantage, and the upper bound is for sampling.}
    \label{fig:several_bins_2}
\end{figure}

The apparently contradictory result shown in Fig. \ref{fig:several_bins_2} is related to the distinction between {\sl accuracy} and {\sl precision}. {\sl Accuracy} stands for how close is a measurement to the exact value of a quantity, and {\sl precision} encodes the dispersion of different measurements. Amplitude Estimation is an algorithm to increase the precision of a measurement with respect to the number of samples, but it does not provide any further information regarding the accuracy. Amplitude Estimation for the binary algorithm reflects the expected tendency for the increase in precision, but comes with very poor results in accuracy. The unary algorithm grows slower in terms of precision, but maintains more accurate results. This decrease in precision might lead to losing the quantum advantage provided by AE in the presence of significant error. We study further in App. \ref{sec:ap_iae} the limit of AE iterations that can be performed given the error rates of the quantum device while still maintaining quantum advantage for the unary representation.

\end{document}